\documentclass{article}
\usepackage[utf8]{inputenc}
\usepackage[margin=0.85in]{geometry}
\usepackage{subcaption}
\usepackage{adjustbox}
\usepackage[percent]{overpic}
\usepackage{amsfonts}
\usepackage{amsmath}
\usepackage[utf8]{inputenc}

\usepackage{authblk}
\usepackage{color}

\newcommand*{\myfont}{\fontfamily{phv}\fontseries{ul}\selectfont}

\title{Optimizing the Human Learnability of Abstract Network Representations}

\author[a]{William Qian}
\author[b,c]{Christopher W. Lynn}
\author[d]{Andrei A. Klishin}
\author[d]{Jennifer Stiso}
\author[e]{Nicolas H. Christianson}
\author[a,d,f,g,h,i]{Dani S. Bassett}

\affil[a]{Department of Physics \& Astronomy, College of Arts \& Sciences, University of Pennsylvania, Philadelphia, PA 19104}
\affil[b]{Initiative for the Theoretical Sciences, Graduate Center, City University of New York, New York, NY 10016}
\affil[c]{Joseph Henry Laboratories of Physics and Lewis–Sigler Institute for Integrative Genomics, Princeton University, Princeton, NJ 08544, USA}
\affil[d]{Department of Bioengineering,
School of Engineering \& Applied Science, University of Pennsylvania, Philadelphia, PA 19104}
\affil[e]{Department of Computing and Mathematical Sciences, Caltech, Pasadena, CA 91125}
\affil[f]{Department of Electrical \& Systems Engineering, School of
Engineering \& Applied Science, University of Pennsylvania, Philadelphia, PA 19104}
\affil[g]{Department of Neurology, Perelman School of Medicine, University of
Pennsylvania, Philadelphia, PA 19104}
\affil[h]{Department of Psychiatry, Perelman School of Medicine, University of Pennsylvania, Philadelphia, PA 19104}
\affil[i]{Santa Fe Institute, Santa Fe, NM 87501}

\begin{document}

\maketitle

\begin{abstract}
Precisely how humans process relational patterns of information in knowledge, language, music, and society is not well understood. Prior work in the field of statistical learning has demonstrated that humans process such information by building internal models of the underlying network structure. However, these mental maps are often inaccurate due to limitations in human information processing. The existence of such limitations raises clear questions: Given a target network that one wishes for a human to learn, what network should one present to the human? Should one simply present the target network as-is, or should one emphasize certain parts of the network to proactively mitigate expected errors in learning? To answer these questions, we study the optimization of network learnability. Evaluating an array of synthetic and real-world networks, we find that learnability is enhanced by reinforcing connections within modules or clusters. In contrast, when networks contain significant core-periphery structure we find that learnability is best optimized by reinforcing peripheral edges between low-degree nodes. Overall, our findings suggest that the accuracy of human network learning can be systematically enhanced by targeted emphasis and de-emphasis of prescribed sectors of information. 
\end{abstract}

\section*{Introduction}
From a young age, humans demonstrate the capacity to learn the relationships between concepts \cite{Saffran_Aslin_Newport_1996,Fiser_Aslin_2002,Perani_Saccuman_Scifo_Anwander_Spada_Baldoli_Poloniato_Lohmann_Friederici_2011}. During the learning process, humans are exposed to discrete chunks of information that combine and interconnect to form cognitive maps that can be represented as complex networks \cite{sizemore2018knowledge,karuza2016local,engelthaler2017feature,solomon2019implementing,peer2021structuring,behrens2018what}. These chunks of information often appear in a natural sequential order, such as words in language, notes in music, and abstract concepts in stories and classroom lectures \cite{Sole_Corominas-Murtra_Valverde_Steels_2010, Liu_Tse_Small_2010, Miller_Fellbaum_1991, herring_paolillo_ramos-vielba_kouper_wright_stoerger_scheidt_clark_2007, Peretz_Gosselin_Belin_Zatorre_Plailly_Tillmann_2009}.  Further, these sequences are encoded in the brain as networks, with links between items reflecting observed transitions (see \cite{schapiro2013neural,schapiro2016statistical,Tompson_Kahn_Falk_Vettel_Bassett_2020, Stiso2021} for empirical studies and \cite{rueckemann2021grid} for a recent review). Broadly, the fact that many different types of information exhibit temporal order (and therefore network structure) motivates investigations into the processes that underlie the human learning of transition networks \cite{peer2021structuring, rueckemann2021grid, Lynn_Bassett_2020}.

To understand the network learning process, recent studies have investigated how humans internally construct abstract representations of associations \cite{Lynn_Kahn_Nyema_Bassett_2020,momennejad2017successor,stachenfeld2017hippocampus}. Using a variety of approaches, from computational models to artificial neural networks, such studies have consistently found that the mind builds network representations by integrating information over time. Such integration enables humans to compress exact sequences of experienced events into broader, but less precise, representations of context \cite{Dayan1993}. These mental representations allow learners to make better generalizations about new information, at the cost of accuracy \cite{momennejad2017successor}. Here we focus on one particular modeling approach that accounts for the temporal integration and inaccuracies inherent in human learning. In particular, we build upon a maximum entropy model which posits that the mind learns a network representation of the world in a manner guided by a tradeoff between accuracy and complexity\cite{Lynn_Kahn_Nyema_Bassett_2020,Lynn_Papadopoulos_Kahn_Bassett_2020}. Specifically, in order to conserve mental resources, humans will tend to reduce the complexity of their representations at the cost of accuracy by introducing errors into the learning process.

While inaccuracies in human learning can aid flexibility across contexts, they present fundamental obstacles for the human comprehension of transition networks. Thus, a clear question emerges: What strategies should be employed to most effectively communicate the structure of a network to an inaccurate human learner? Prior studies of animal communication and behavior have demonstrated the utility of exaggerating the presentation of certain signals to receivers in offsetting erroneous information processing \cite{Wiley_2015, real_1994}. Similarly, one could imagine that, by emphasizing some features of a network over others, one may be able to correct for errors in human learning. Such an approach of targeted modulation of emphasis may be helpful not only in learning a whole network, but also in optimally learning particularly challenging parts of a network. In fact, humans show consistent difficulties in learning certain motifs in networks, such as the connections between modules \cite{karuza2017process,kahn2018network,Lynn_Kahn_Nyema_Bassett_2020,Tompson_Kahn_Falk_Vettel_Bassett_2019}. Taken together, these observations suggest that disproportionately weighting specific network features that are difficult to learn may facilitate human network learning.

\section*{Mathematical Methods}
To study the optimization of network learnability, we first require a model describing how humans learn networks. Following Ref. \cite{Lynn_Kahn_Nyema_Bassett_2020}, we write down an analytic model that captures a wide range of human behaviors observed in network learning experiments. Specifically, we consider a transition probability matrix $A \in \mathbb{R}^{n\times n}$ that describes random walks on the observed network, where $A_{ij}$ represents the probability of a transition to node $j$ when starting from node $i$. Then, after observing a sufficient number of random walks, the human's internal representation of the network converges to the following analytic form:
\begin{equation}
    f(A) = (1 - e^{-\beta})A(I - e^{-\beta}A)^{-1},
\end{equation}
where the single parameter $\beta$ reflects the accuracy of the human learner. In the limit $\beta \rightarrow 0$, the learned network structure $f(A)$ is a fully connected network with uniform edge weights, and hence bears no resemblance to the actual network $A$. In contrast, in the limit $\beta \rightarrow \infty$, the learning process is free of errors and the learned network structure $f(A)$ is an exact replica of the actual network $A$.

The following question then arises: Given a transition network $A$ and a human learner with accuracy $\beta$, what is the optimal input transition network $A_{\text{in}} = A^*$ that, when presented to a human learner, results in a perceived network structure $f(A_{\text{in}})$ that most closely matches the true structure $A$? In general, there is no reason to presume that it is optimal to present the learner with the true network (such that $A^* = A$). Indeed, teaching and other forms of communication often involve the purposeful emphasis or exaggeration of some pieces of information over others \cite{Bakir_Herring_Miller_Robinson_2019, Bloedel_Segal_2018, Boleslavsky_Kim_2017}. Thus, it is possible that modulating emphasis on certain network features, in a precise and targeted manner that serves to counteract natural biases or expected errors, might enhance learnability. Moreover, optimal emphasis strategies are likely to vary from person to person, depending critically upon the accuracy $\beta$ of the human's learning process.

\begin{figure*}[t]
\begin{subfigure}[t]{0.015\textwidth}
  \vspace{-5.02cm}
  {\myfont \Large A}
\end{subfigure}
  \adjustbox{minipage=1em}{\label{sfig:testa}}%
  \begin{subfigure}[t]{\dimexpr.325\linewidth}
  \centering
  \vspace{-4.7cm}
  
  \begin{overpic}[width=0.9\textwidth,tics=10]{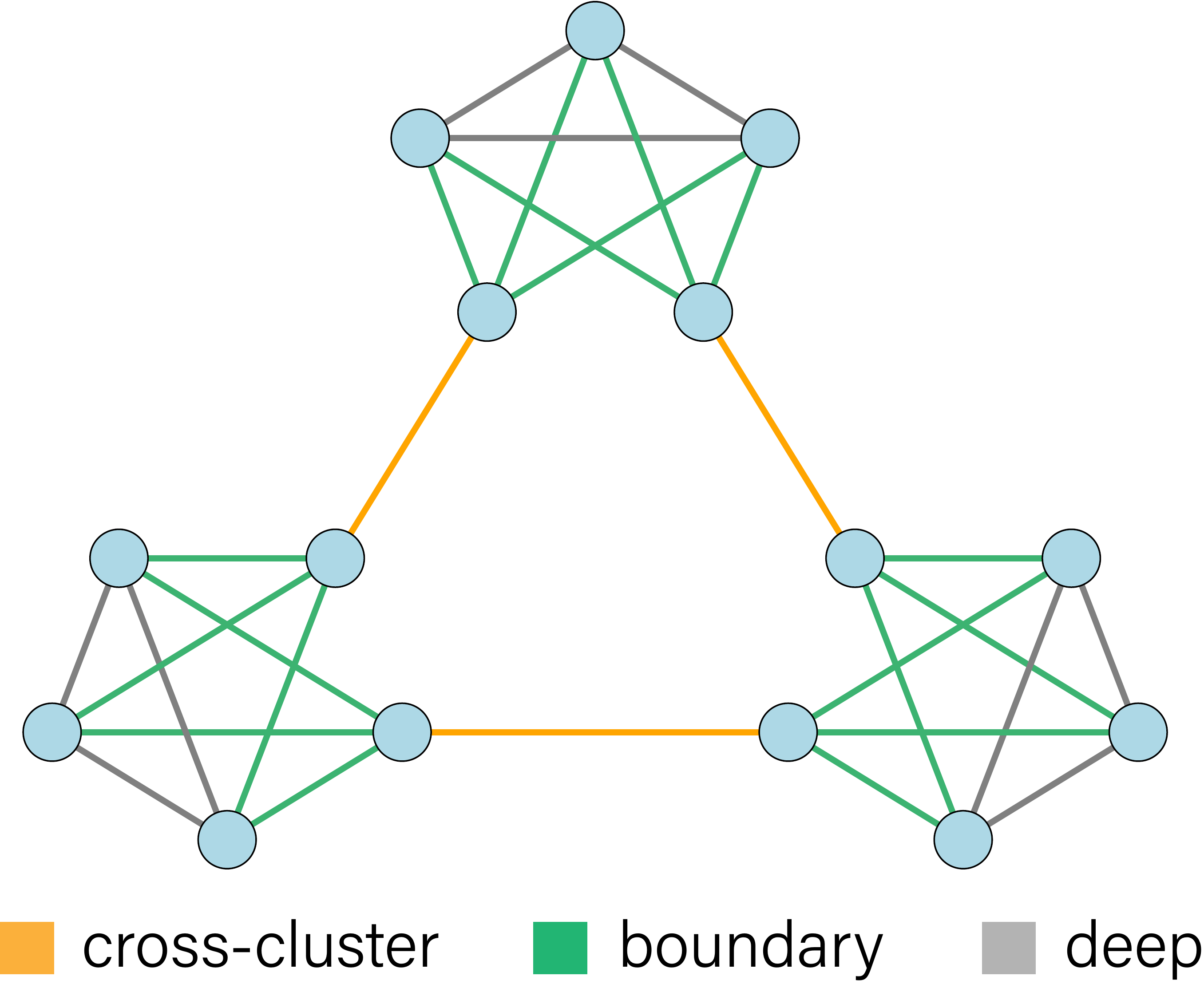}
  \put (45,14) {$\lambda_{cc}$}
  \put (68,44) {$\lambda_{b}$}
  \end{overpic}
  \end{subfigure}%
  \begin{subfigure}[t]{0.013\textwidth}
  \vspace{-5.02cm}
  {\myfont \Large C}
\end{subfigure}
  \adjustbox{minipage=1em}{\label{sfig:testb}}%
  \begin{subfigure}[t]{\dimexpr.345\linewidth}
  \centering
  \begin{overpic}[trim = 70 0 0 0, clip, scale=.43, tics=10]{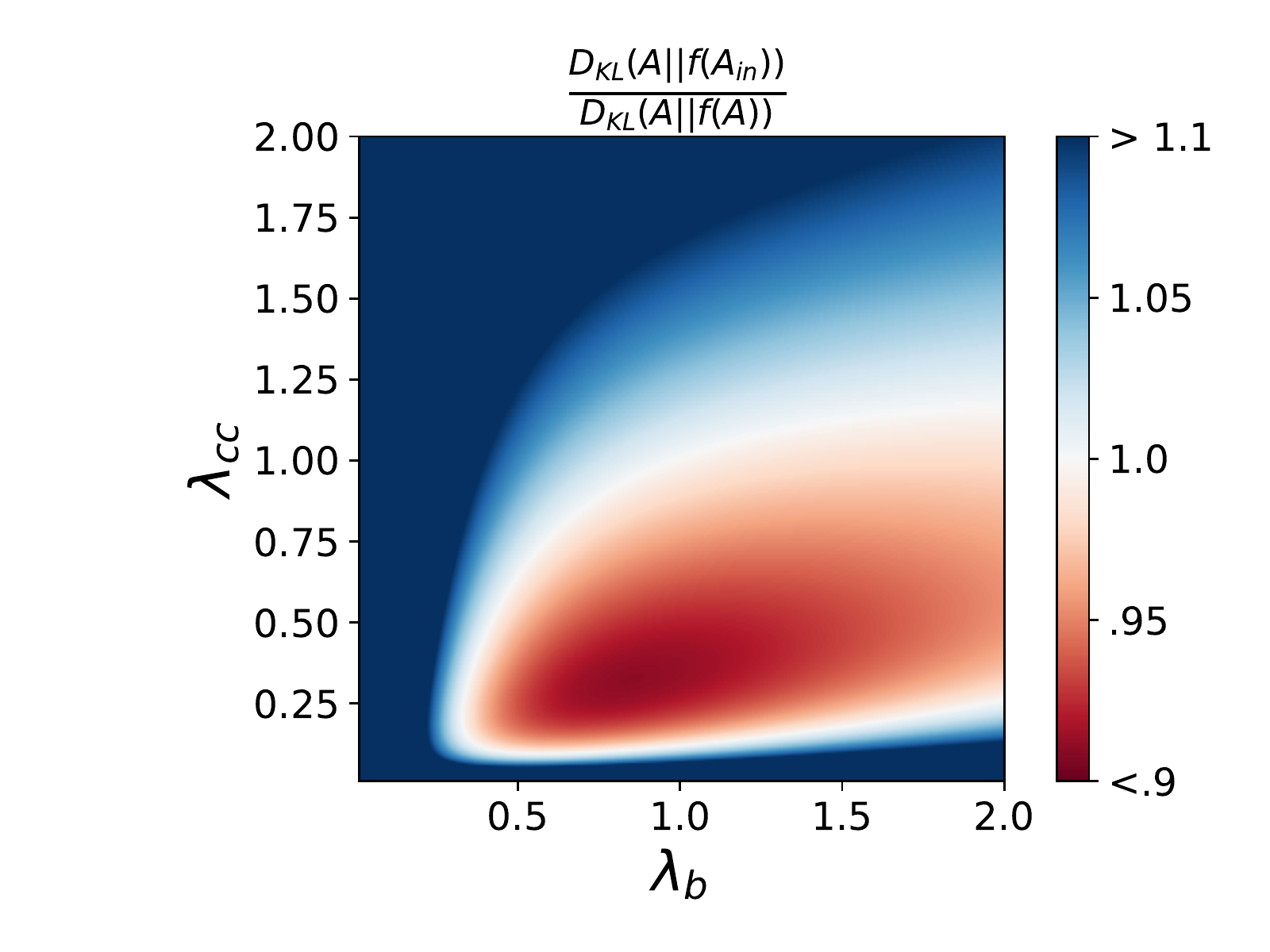}
  \put (54, 68) {\color{green} $\beta = 0.3$}
  \end{overpic}
  \end{subfigure}
  \begin{subfigure}[t]{0.015\textwidth}
  \vspace{-5.02cm}
  {\myfont \Large E}
\end{subfigure}
  \adjustbox{minipage=1em}{\label{sfig:testa}}%
  \begin{subfigure}[t]{\dimexpr.25\linewidth}
  \centering 
    \vspace{-4.65cm}
      \includegraphics[trim = 60 0 0 0, scale=.39]{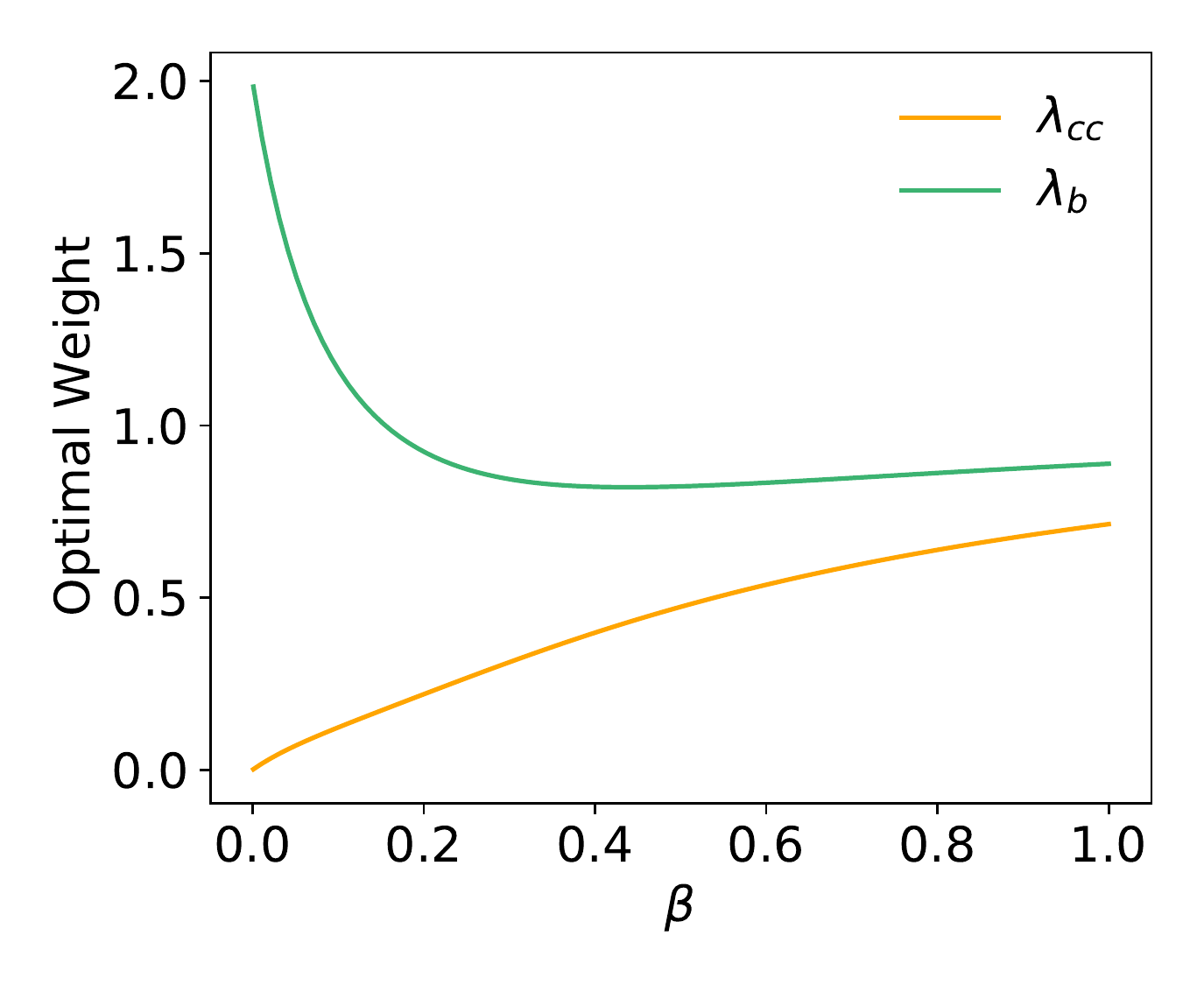}
  \end{subfigure}%
  \\
  \vspace{-1.3cm}
  
  \begin{subfigure}[t]{0.015\textwidth}
  \vspace{.5cm}
  {\myfont \Large B}
\end{subfigure}
  \adjustbox{minipage=1em}{\label{sfig:testa}}%
  \begin{subfigure}[t]{\dimexpr.325\linewidth}
  \vspace{1em}
  \centering
  
  \begin{overpic}[trim = 70 0 0 0, clip, scale=.43, tics=10]{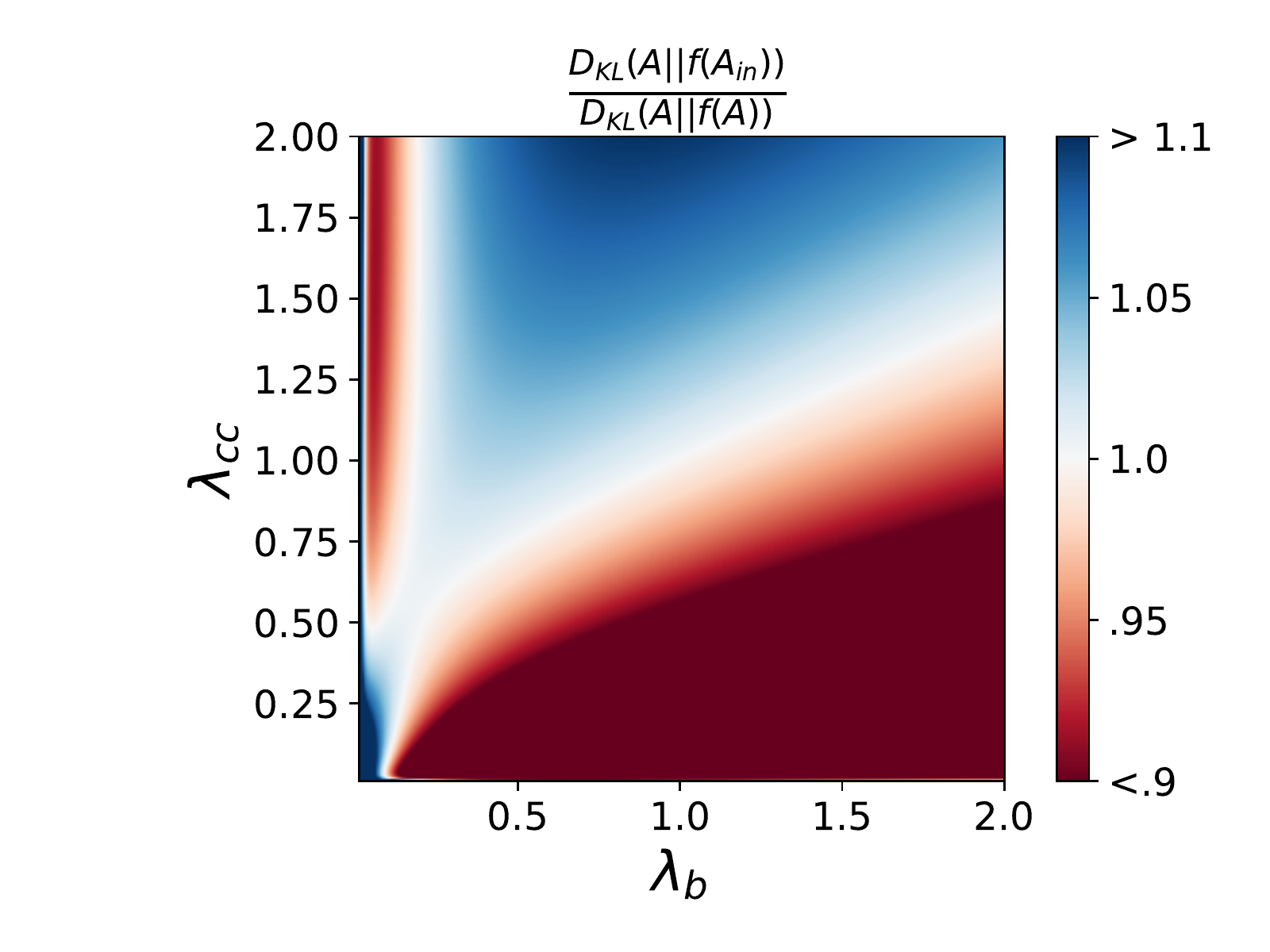}
  \put (50, 68) {\color{green} $\beta = 0.05$}
  \end{overpic}
  \end{subfigure}%
  \begin{subfigure}[t]{0.013\textwidth}
  \vspace{.5cm}
 {\myfont \Large D}
\end{subfigure}
\adjustbox{minipage=1em}{\label{sfig:testa}}%
  \begin{subfigure}[t]{\dimexpr.345\linewidth}
  \vspace{1em}
  \centering
    \begin{overpic}[trim = 70 0 0 0, clip, scale=.43, tics=10]{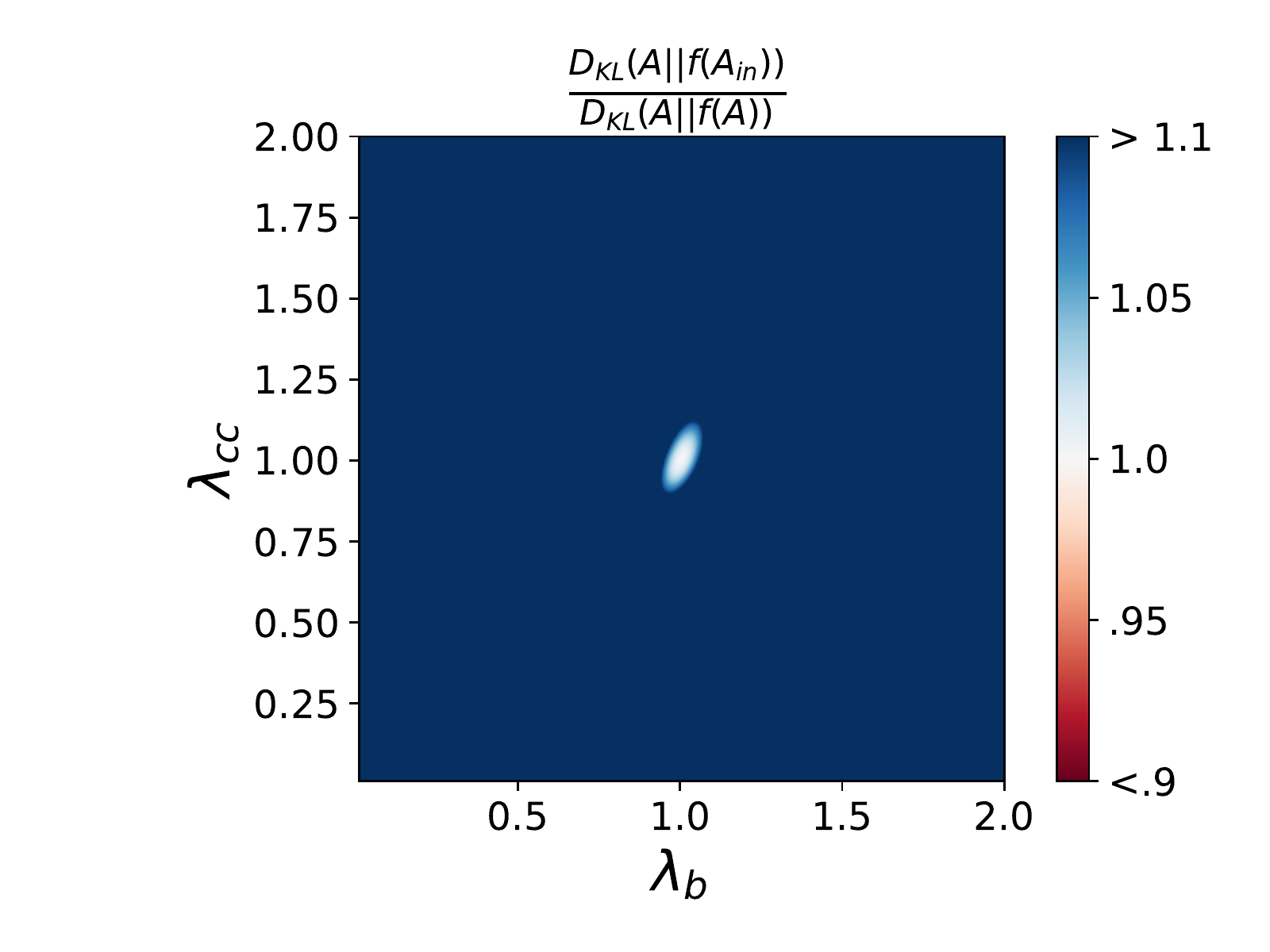}
    \put (58, 68) {\color{green} $\beta = 5$}
  \end{overpic}
  \end{subfigure}%
  \begin{subfigure}[t]{0.015\textwidth}
  \vspace{.5cm}
 {\myfont \Large F}
\end{subfigure}
\adjustbox{minipage=1em}{\label{sfig:testa}}%
  \begin{subfigure}[t]{\dimexpr.25\linewidth}
  \vspace{2.5em}
  \centering
      \includegraphics[trim = 50 0 0 0, scale=.39]{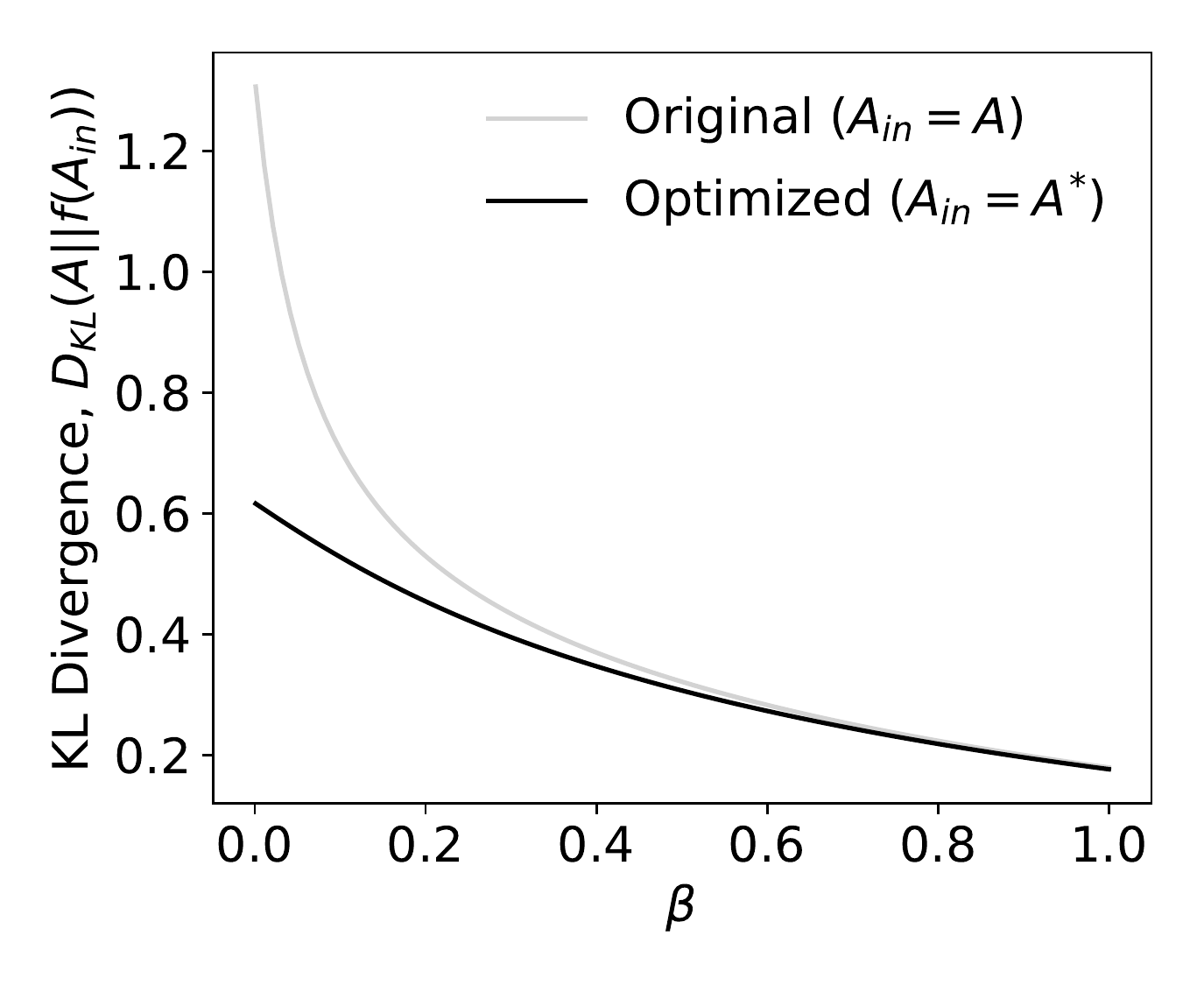}
  \end{subfigure}%
      \caption{\textbf{Optimizing the learnability of a modular graph.} \emph{(A)} A modular graph with $15$ nodes, each with degree $k_{i} = 4$, resulting in $30$ edges. \emph{(B-D)} Here we show the Kullback-Leibler divergence ratio (less than $1$ indicates enhanced learnability) across a section of the $\lambda_{cc}$, $\lambda_{b}$ parameter space, for different values of $\beta$. \emph{(B)} 
      Results for $\beta = 0.05$, corresponding roughly to the median accuracy of human learners in prior studies \cite{Lynn_Kahn_Nyema_Bassett_2020}. \emph{(c)} Results for $\beta = 0.3$, corresponding to the mean accuracy of human learners in prior studies \cite{Lynn_Kahn_Nyema_Bassett_2020}. \emph{(D)} Results for $\beta = 5$, corresponding to an exceptionally accurate network learner. \emph{(E)} The optimal edge weights $\lambda_{cc}$ and $\lambda_{b}$ for $0 < \beta < 1$. \emph{(F)} The Kullback-Leibler divergence between the learned network and the true network for different values of $\beta$, both with and without input network optimization.}
  \label{fig:1}
\end{figure*}

\begin{figure*}[h]
\begin{subfigure}[t]{0.015\textwidth}
  \vspace{-2.5cm}
  {\myfont \Large A}
\end{subfigure}
\begin{subfigure}{\dimexpr.44\linewidth}
    \centering
    \begin{overpic}[width=.95\textwidth,tics=10]{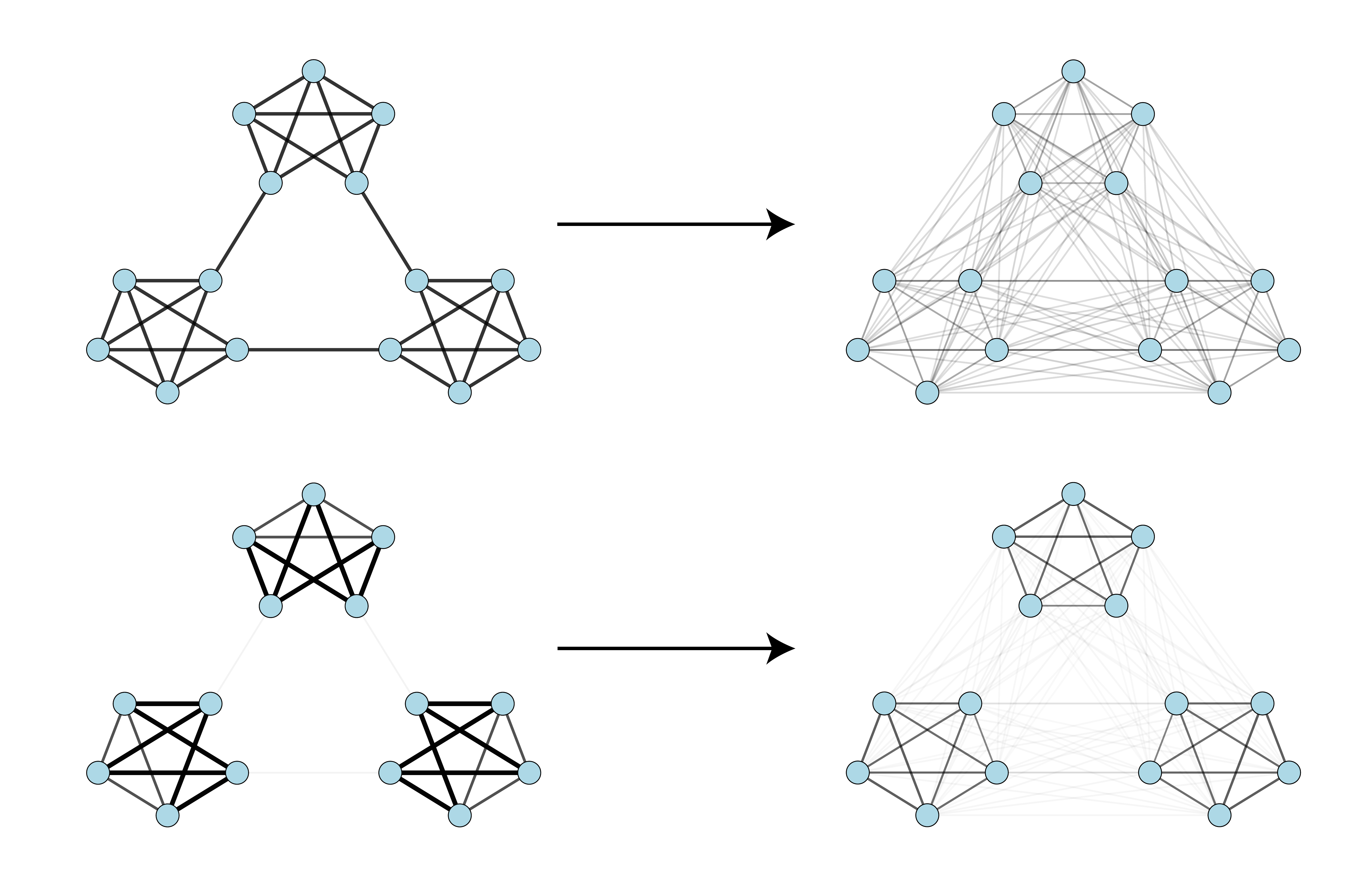}
 \put (33,50) {\Large$\displaystyle A$}
 \put (33, 18.5) {\Large$\displaystyle A^{*}$}
 \put (92, 50) {\Large$\displaystyle f(A)$}
 \put (92, 18.5) {\Large$\displaystyle f(A^{*})$}
\end{overpic}
\end{subfigure}
\hspace{1em}
\begin{subfigure}[t]{0.015\textwidth}
  \vspace{-2.5cm}
  {\myfont \Large B}
\end{subfigure}
\begin{subfigure}{\dimexpr.44\linewidth}
    \centering
    \begin{overpic}[width=.95\textwidth,tics=10]{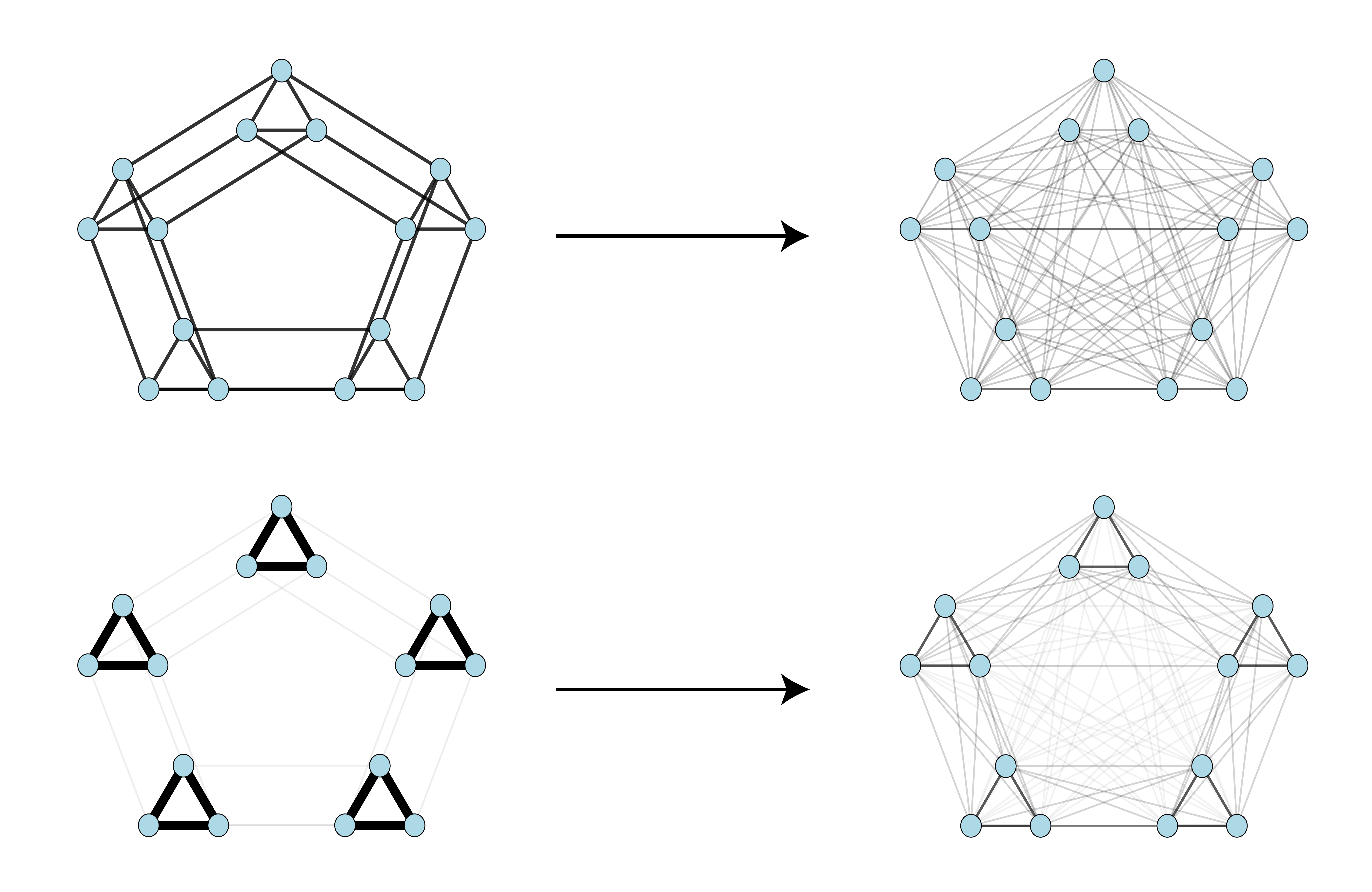}
 \put (36,50) {\Large$\displaystyle A$}
 \put (36, 18.5) {\Large$\displaystyle A^{*}$}
 \put (97, 50) {\Large$\displaystyle f(A)$}
 \put (97, 18.5) {\Large$\displaystyle f(A^{*})$}
\end{overpic}
\end{subfigure}
    \caption{\textbf{Optimal emphasis modulation of the modular and lattice networks.} Here we show the learned networks resulting from human learning of the modular and lattice networks, respectively \emph{(A, B)} (top), as well as from the modular and lattice networks optimized for learnability \emph{(A, B)} (bottom). Optimized and learned networks were both computed at $\beta = 0.05$. Edge thickness indicates transition probabilities.}
    \label{fig:2}
\end{figure*}
One natural approach to answering this generic question is to find the input matrix $A_{\text{in}}$ such that the learned representation $f(A_{\text{in}})$ is equal to the target network $A$. From Equation 1, one can derive for a given $\beta$ an analytical form for the input $A_{\text{in}} = A^{*}$ such that $f(A^{*}) = A$ holds exactly. While this approach is mathematically elegant, it has limited application to real-world scenarios because the resulting $A^{*}$ often contains negative entries and is hence an ill-defined transition matrix. To overcome this hurdle, one can instead characterize how a perceived network structure $f(A_{\text{in}})$ diverges from some true structure $A$ using the Kullback-Leibler divergence $D_{KL}(A||f(A_{\text{in}}))$. To determine the optimal input $A_{\text{in}} = A^{*}$ such that $A^{*}$ is a well-defined transition matrix, we determine a weighted network with adjacency matrix $G_{\text{in}}$ such that the corresponding transition matrix $A_{\text{in}}$ minimizes the Kullback-Leibler divergence $D_{KL}(A||f(A_{\text{in}}))$ between the learned structure $f(A_{\text{in}})$ and the true structure $A$. Practically, we implement this strategy using dual annealing, which is a powerful and common method for bounded optimization \cite{Tsallis_1988, Tsallis_Stariolo_1996, Xiang_Sun_Fan_Gong_1997}.

For Large input networks with many edges, this optimization process can become computationally unwieldy. To address this issue, we only consider input networks $G_{\text{in}}$ that respect the symmetries of $A$: all structurally unique edges in $A$ must have the same edge weight in $G_{\text{in}}$. Further, we only consider the inclusion of edges in $G_{\text{in}}$ when they have a counterpart in $A$ with nonzero weight. In this manner, the network optimization process can be parameterized by a significantly smaller number of trainable values for networks with a high degree of symmetry. 

To investigate possible strategies for enhancing transition network learnability, we apply the optimization method to two transition networks: a modular network and a lattice network. Both of these networks have $15$ vertices, and share the property that every node has degree $4$. Importantly, previous human experiments were able to directly estimate the accuracy parameter $\beta$ of human learning in these networks \cite{Lynn_Kahn_Nyema_Bassett_2020}. Next, to explore the optimization of learnability in asymmetric networks with nonuniform transition probabilities, we consider the optimization of learnability for networks constructed from generative network models. Lastly, to probe how real-world information networks ought to be designed, we investigate how learnability can be maximized for semantic networks extracted from college mathematics textbooks. 

In performing these numerical experiments, we are guided by several hypotheses. Specifically, in considering prior work demonstrating the efficacy of exaggeration in animal communication \cite{Wiley_2015, real_1994}, we predict that strategic modulation of emphasis will significantly improve learnability in both synthetic and real-world information networks. Furthermore, in view of prior work demonstrating that highly-clustered networks are more learnable than lattice or random networks \cite{Lynn_Papadopoulos_Kahn_Bassett_2020}, we hypothesize that optimal emphasis modulation strategies will reinforce connections within clusters and de-emphasize connections across clusters. And finally, given the human tendency to attend to salient information, we hypothesize that core-periphery networks will be best learned by emphasizing edges in the network periphery which would otherwise be less easily learned. Taken together, our numerical experiments aim to assess whether and how the accuracy of human network learning can be systematically enhanced by targeted emphasis and de-emphasis of network features. 

\section*{Results}

\subsection*{Optimizing the learnability of graph exemplars} 

\paragraph{The modular graph exemplar.} We begin by studying the learning optimization of the modular graph shown in Fig. \ref{fig:1}\emph{A}, which has previously been used in human learning studies \cite{schapiro2013neural,karuza2017process,kahn2018network,Tompson_Kahn_Falk_Vettel_Bassett_2019}. In this graph, there are only $3$ structurally unique edges: cross-cluster edges (orange), boundary edges that are adjacent to cross-cluster edges (green), and edges deep within modules (grey). Thus, the structure of $A_{\text{in}}$ can be determined by two free parameters $\lambda _{cc}$ and $\lambda_{b}$, representing the weights of cross-cluster and boundary edges in $G_{\text{in}}$, respectively, relative to the weight of deep edges in $G_{\text{in}}$. Note that one free parameter has been removed due to the constraint that $A_{\text{in}}$ is normalized such that all rows sum to 1.  

To illustrate the parameter regimes where certain weighting combinations of $\lambda_{cc}$ and $\lambda_{b}$ are effective in enhancing learnability, we first computed the ratio $\frac{D_{KL}(A||f(A_{\text{in}}))}{D_{KL}(A||f(A))}$ over the parameter space $0 \leq \lambda_{cc} \leq 2$, $0 \leq \lambda_{b} \leq 2$, for three different values of $\beta$ (Figs. \ref{fig:1}B-D). This ratio characterizes the learnability that can be achieved by targeted modulation of emphasis in the network. Specifically, a ratio of less than $1$ would indicate that the emphasized network improves learnability over the true network. 

Interestingly, at low values of $\beta$ (Fig.\ref{fig:1}\emph{B}, $\beta = 0.05$), when the learning process is highly inaccurate, there are two regimes in which an emphasized network structure improves learnability: one that heavily de-emphasizes boundary edges, and one that moderately de-emphasizes cross-cluster edges. For intermediate values of $\beta$ (such as $\beta = 0.3$ in Fig. 1C), the two optimal regimes combine into one. As the learning accuracy increases further (Fig. 1D, $\beta$ = 5), the one optimal regime decreases in size and converges to the true network structure. Thus, for extremely precise learners, the only reasonable network structure to learn would be the true network structure, corresponding to $\lambda_{cc} = \lambda_{b} = 1$ (Fig.\ref{fig:1}\emph{D}, $\beta = 5$). 

To assess the precise values of edge weights that lead to optimal learning of the modular graph, we minimize $D_{KL}(A||f(A_{\text{in}}))$ with respect to $\lambda_{cc}$ and $\lambda_{b}$ at different values of $\beta$ (Fig. \ref{fig:1}\emph{E}). We find that cross-cluster edges are always de-emphasized, whereas boundary edges are over-emphasized for inaccurate learners but de-emphasized for the average human learner ($\beta \approx 0.3$). We present a graphical depiction of the optimal input network and the resulting learned structure for the modular graph at $\beta = 0.05$ in Fig. \ref{fig:2}\emph{A}. 

\begin{figure*}[t]
\begin{subfigure}[t]{0.015\textwidth}
  \vspace{-5.05cm}
  {\myfont \Large A}
\end{subfigure}
  \adjustbox{minipage=1em}{\label{sfig:testa}}%
  \begin{subfigure}[t]{\dimexpr.3\linewidth}
  \centering
  \vspace{-5.3cm}
  \begin{overpic}[trim = 50 0 0 0, clip, scale=.4, tics=10]{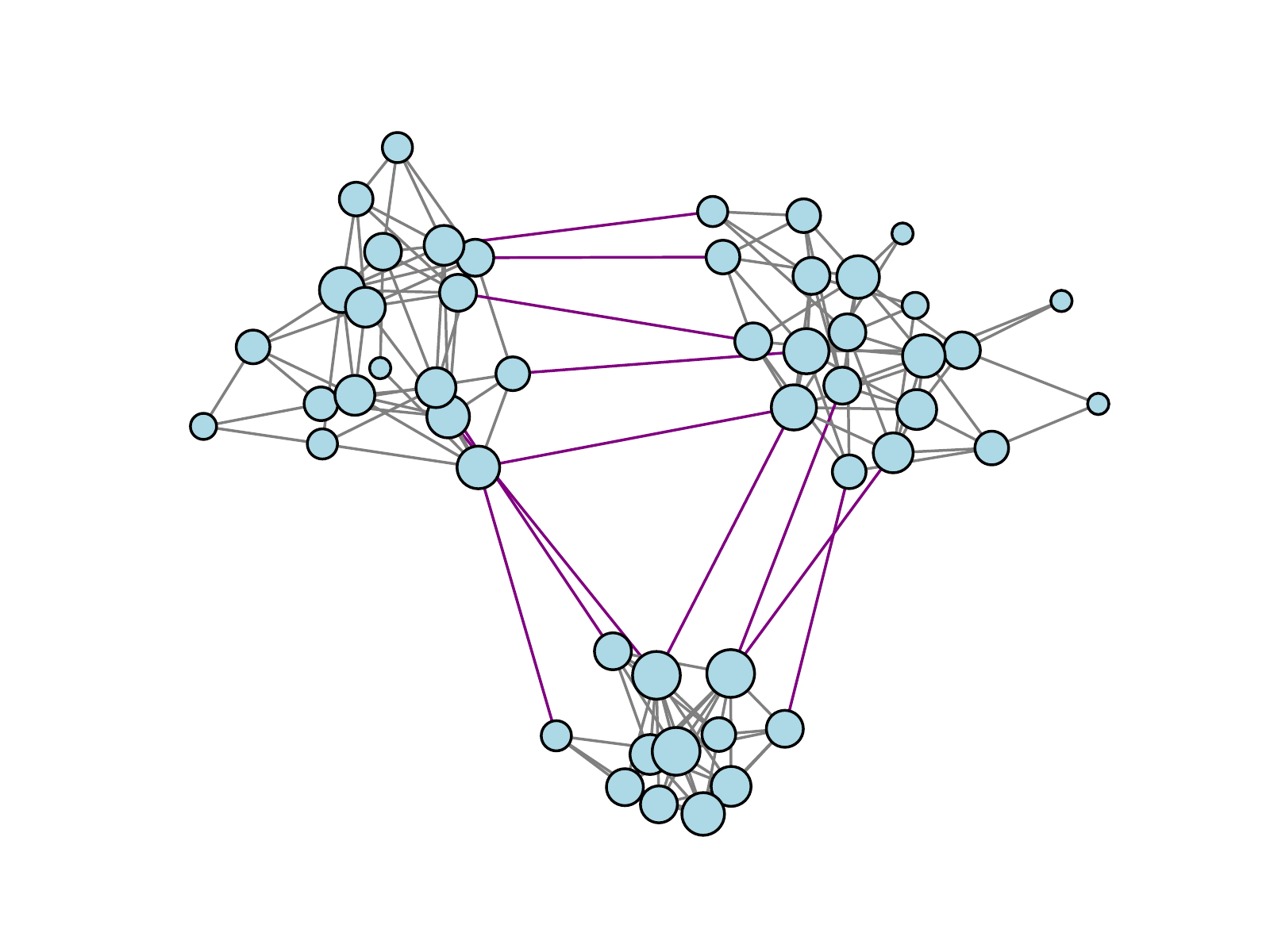}
  \put (39, 2) {\myfont Standard}
  \end{overpic}
  \end{subfigure}%
  \begin{subfigure}[t]{0.015\textwidth}
  \vspace{-5.05cm}
  {\myfont \Large C}
\end{subfigure}
  \adjustbox{minipage=1em}{\label{sfig:testb}}%
  \begin{subfigure}[t]{\dimexpr.32\linewidth}
  \centering
        \includegraphics[trim = 15 15 0 0, clip, scale=.427]{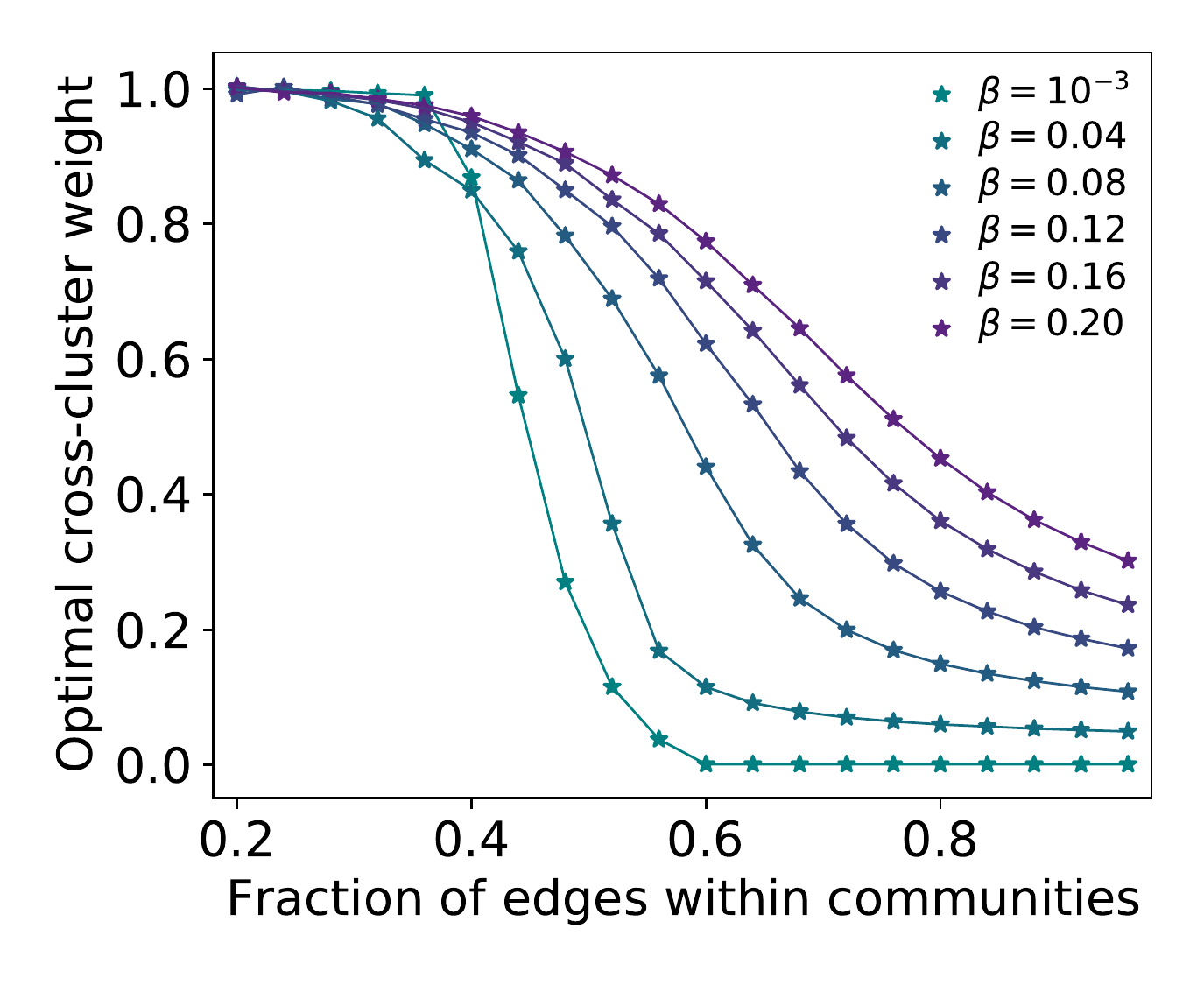}
  \end{subfigure}
  \begin{subfigure}[t]{0.015\textwidth}
  \vspace{-5.05cm}
  {\myfont \Large E}
\end{subfigure}
  \adjustbox{minipage=1em}{\label{sfig:testa}}%
  \begin{subfigure}[t]{\dimexpr.32\linewidth}
  \centering 
      \includegraphics[trim =20 15 0 0, scale=.43]{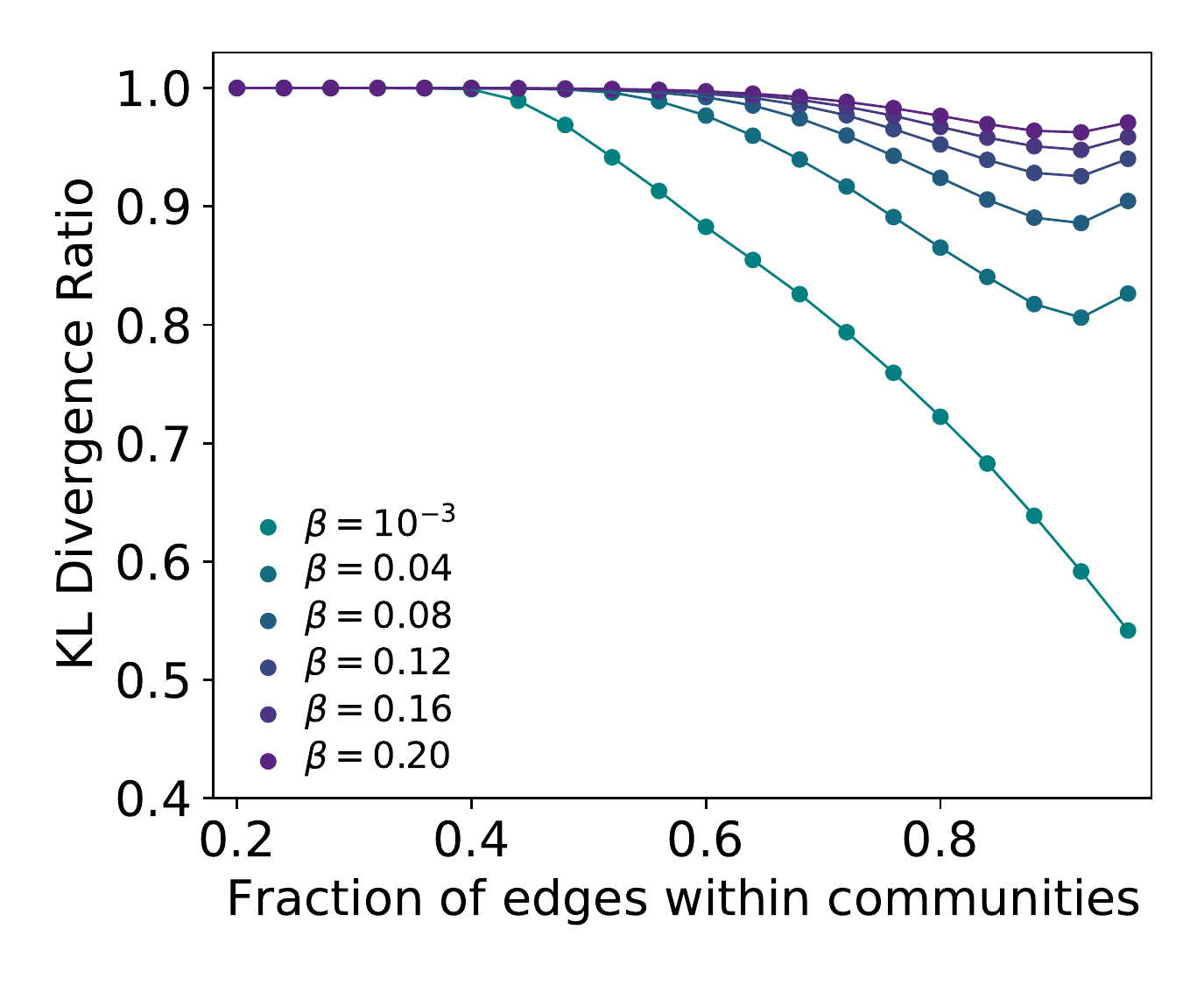}
  \end{subfigure}%
  \\
  \vspace{-1.3cm}
  
  \begin{subfigure}[t]{0.015\textwidth}
  \vspace{.5cm}
  {\myfont \Large B}
\end{subfigure}
  \adjustbox{minipage=1em}{\label{sfig:testa}}%
  \begin{subfigure}[t]{\dimexpr.3\linewidth}
  \vspace{1em}
  \centering
  \begin{overpic}[trim = 60 0 0 0, clip, scale=.41, tics=10]{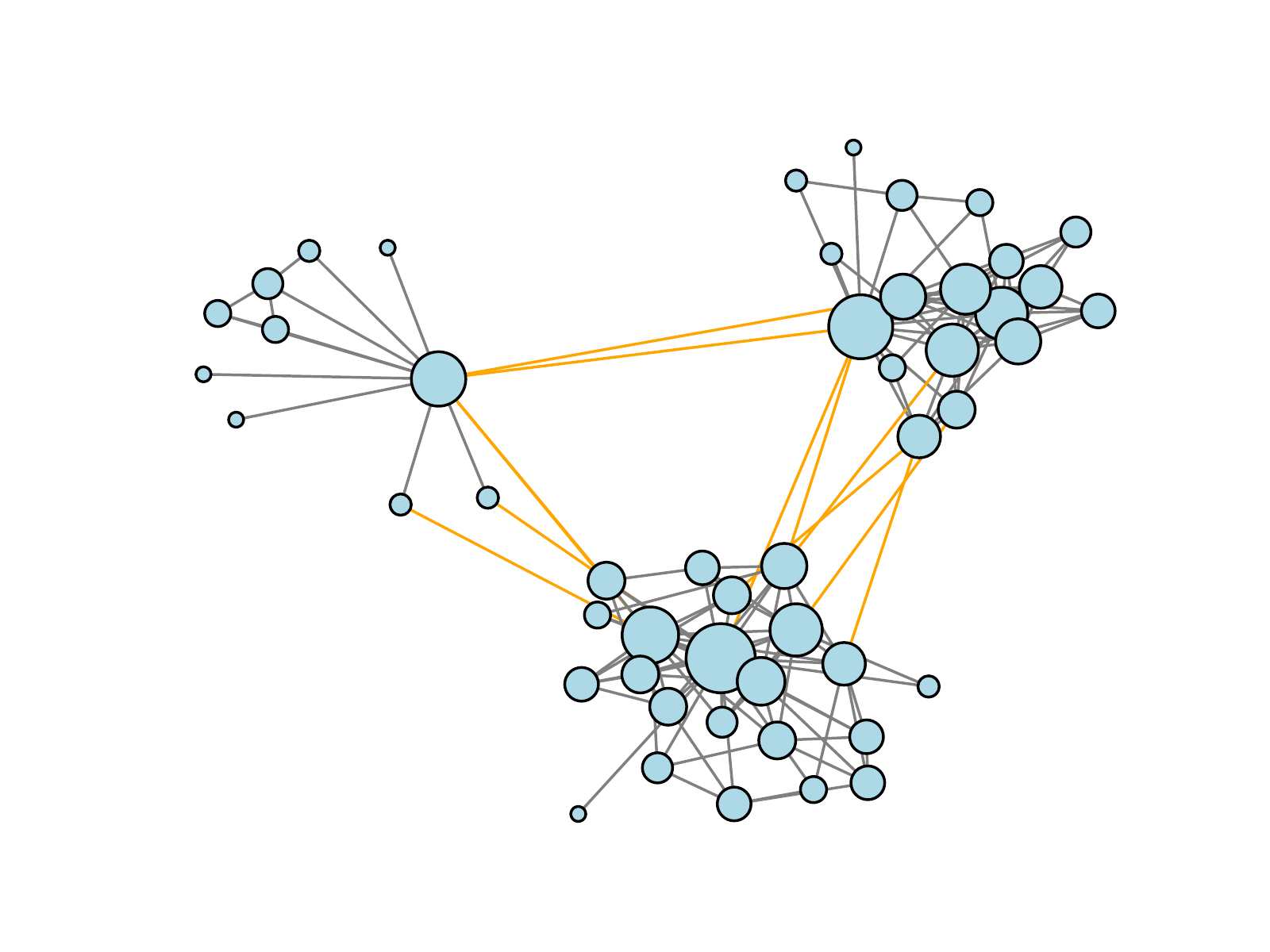}
  \put (30, 4) {\myfont Degree-corrected}
  \end{overpic}
  \end{subfigure}%
  \begin{subfigure}[t]{0.015\textwidth}
  \vspace{.5cm}
 {\myfont \Large D}
\end{subfigure}
\adjustbox{minipage=1em}{\label{sfig:testa}}%
  \begin{subfigure}[t]{\dimexpr.32\linewidth}
  \vspace{2.5em}
  \centering
      \includegraphics[trim = 15 0 0 0, clip, scale=.43]{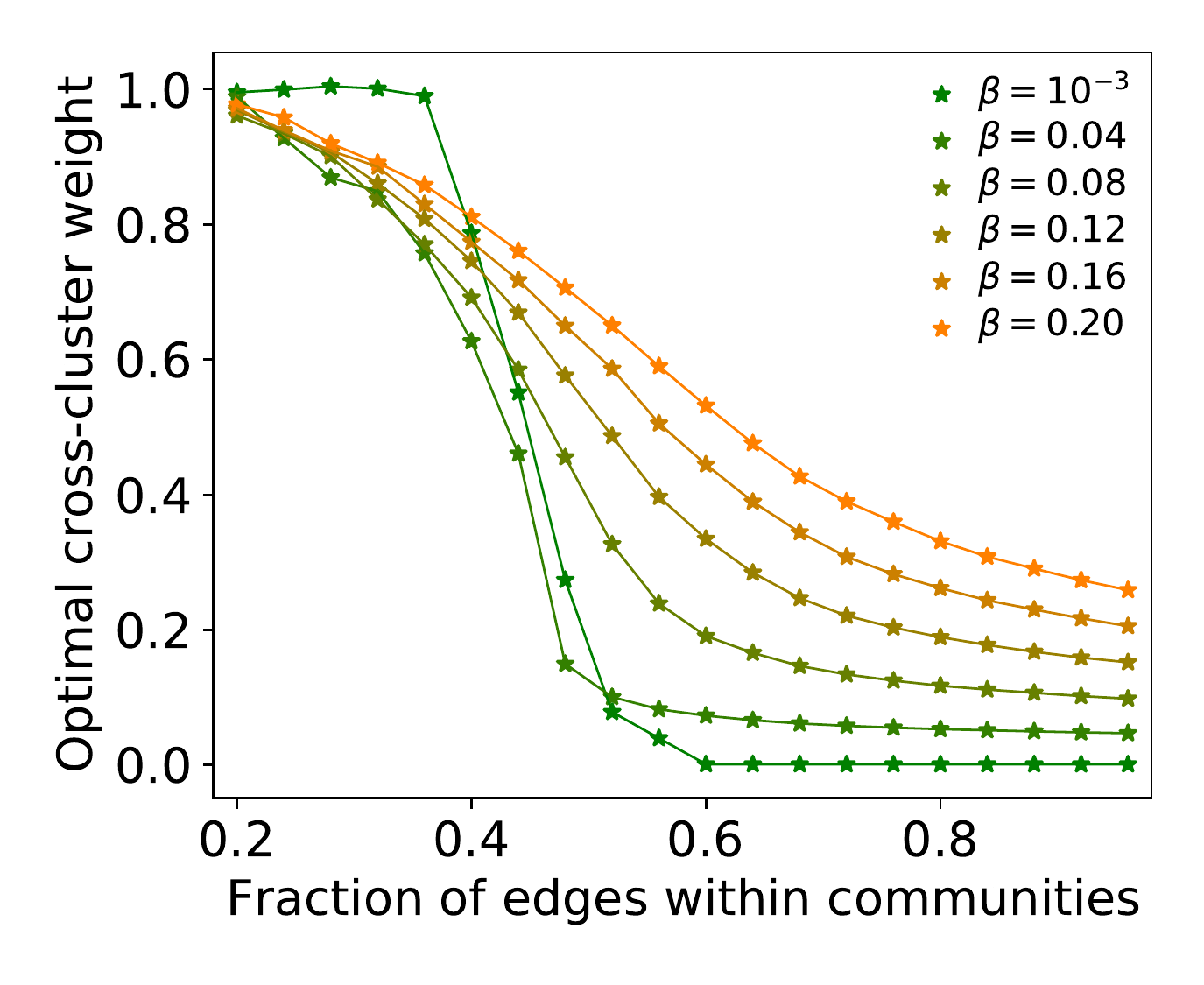}
  \end{subfigure}%
  \begin{subfigure}[t]{0.015\textwidth}
  \vspace{.5cm}
 {\myfont \Large F}
\end{subfigure}
\adjustbox{minipage=1em}{\label{sfig:testa}}%
  \begin{subfigure}[t]{\dimexpr.32\linewidth}
  \vspace{2.5em}
  \centering
      \includegraphics[trim = 20 0 0 0, scale=.43]{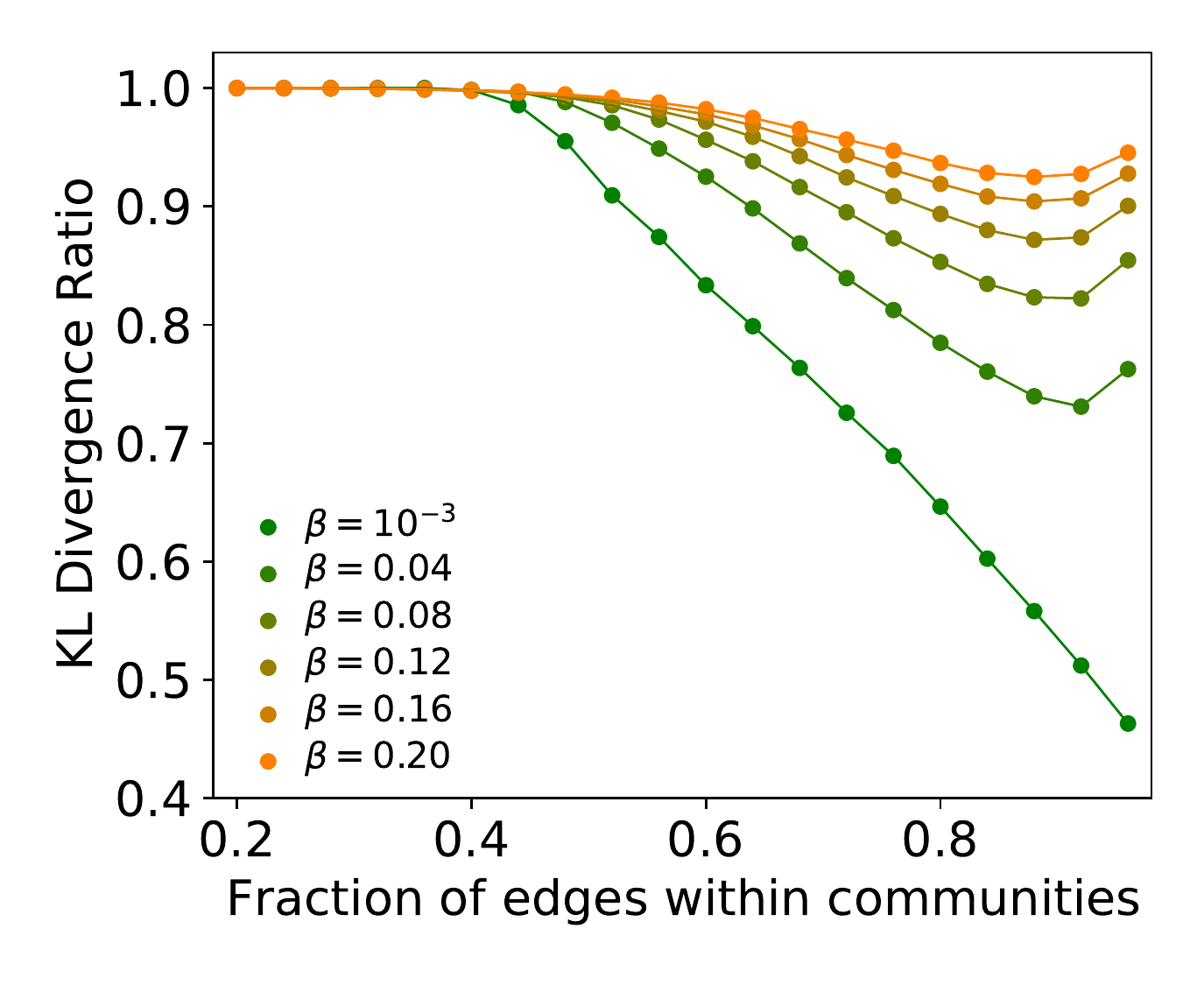}
  \end{subfigure}%
      \caption{\textbf{Optimizing the learnability of synthetic modular networks.} \emph{(A,B)} Examples of a standard stochastic block network and a degree-corrected stochastic block network. Node sizes are proportional to node degrees, with cross-cluster edges shown in purple and orange, respectively. \emph{(C,D)} The optimal cross-cluster edge weight $\lambda_{cc}$ for enhancing learnability versus the fraction $f$ of edges within communities at different values of $\beta$. Results are shown for stochastic block networks and degree-corrected stochastic block networks, respectively. \emph{(E,F)}
      The Kullback-Leibler divergence ratio $\frac{D_{KL}(A||f(A_{\text{in}}))}{D_{KL}(A||f(A))}$ achieved with optimal cross-cluster edge weights at different values of $\beta$. Results are shown for stochastic block networks and degree-corrected stochastic block networks, respectively. The findings reported in panels \emph{(C,D,E,F)} represent results obtained for networks with $N = 200$ nodes, $5$ communities, and an average degree of $\langle k \rangle = 10$. Each curve is an average over the results from $25$ generated networks. }
  \label{fig:3}
\end{figure*}

\paragraph{The lattice graph exemplar.} To understand how optimizing network learnability varies with the topology of the target network, we also study the optimization of learnability of a lattice graph that was examined in human learning studies \cite{Lynn_Kahn_Nyema_Bassett_2020, kahn2018network} (SI Appendix, Fig. S1). While we find qualitative differences in the efficacy of the optimization process for modular and lattice graphs, we find similarly that small cluster-like substructures in the lattice graph are emphasized to maximize learnability, whereas edges between these substructures are de-emphasized. We present a graphical depiction of the optimal input network and the resulting learned structure for the lattice graph at $\beta = 0.05$ in Fig. \ref{fig:2}\emph{B}.

\paragraph{A Sierpiński graph exemplar.} Next, to assess whether the strategy of over-emphasizing edges within clusters and de-emphasizing those between clusters extends to Larger networks with more complex community organization, we also consider a Sierpiński network with hierarchical community structure (SI Appendix, Fig. S2). Consistent with previous findings, we find that de-emphasizing cross-cluster edges at all hierarchical levels of organization is an effective strategy for optimizing learnability. In addition, we observe that cross-cluster edges at the highest level of organization ought to be de-emphasized the most. 

\subsection*{Optimizing the learnability of generated networks}

\paragraph{Stochastic block networks.} Thus far, we have found that the learnability of networks with modular structure is optimized by over-emphasizing the edges within clusters of nodes and de-emphasizing the edges between clusters. However, our analyses have focused on networks with high degrees of structural symmetry. Here, we extend our analysis to randomly-generated networks, studying the optimization of learnability in stochastic block networks.

We consider two classes of stochastic block networks: (i) stochastic block networks in the absence of specific structure-degree correlations, where all cross-cluster edges are equally likely to be included, and all within-cluster edges are equally likely to included, and (ii) degree-corrected stochastic block networks with heterogeneous degree distributions. These classes were chosen to assess whether degree heterogeneity, a common feature of real-world networks \cite{Lerman_Yan_Wu_2016, Barabasi_Albert_1999}, influences the efficacy of strategies for enhancing network learnability in modular networks. In particular, we consider networks with $N = 200$ nodes, $5$ communities, and an average degree of $\langle k \rangle = 10$ (Figs. \ref{fig:3}\emph{A} and \ref{fig:3}\emph{B}; see Materials and Methods for network generation procedures). For a given stochastic block network $G_{SBM}$ with a normalized transition matrix $A_{SBM}$, we parameterize the network presented to learners $G_{\text{in}}$ by a single parameter $\lambda_{cc}$, representing the weight of edges between clusters relative to the weight of edges within clusters. We then compute the cross-cluster weight $\lambda_{cc}$ that optimizes the learnability of the transition network $A_{SBM}$. 

We begin by analyzing the enhancement of learnability for stochastic block networks without structure-degree correlations. In Fig. \ref{fig:3}\emph{C}, we show the optimal cross-cluster weight $\lambda_{cc}$ as a function of the fraction $f$ of edges chosen to be within communities. Importantly, we find that the optimal cross-cluster weight decreases considerably as the modularity of the target stochastic block networks increases. Moreover, for higher $\beta$, we find that optimally emphasized networks maintain more weight on cross-cluster edges, consistent with our earlier analysis of the modular network (Fig. \ref{fig:1}\emph{A}). In addition, increases in the learnability of stochastic block networks (reductions of Kullback-Leibler divergence ratios) are most prominent for values of $f$ above $0.8$, and increase considerably as $\beta$ decreases (Fig. \ref{fig:3}\emph{E}). 

To determine whether the degree heterogeneity impacts the optimization of network learnability, we study degree-corrected stochastic block networks. For such networks, we find that the optimal cross-cluster edge weight $\lambda_{cc}$ decreases faster with increasing $f$ than for standard stochastic block networks (Fig. \ref{fig:3}\emph{D}). Interestingly, for degree-corrected stochastic block networks, the improvements in learnability are significantly Larger than for regular stochastic block networks (Figs. \ref{fig:3}\emph{E-F}). For both types of networks, increases in network learnability are most pronounced for low values of $\beta$, and peak for highly clustered networks with a fraction of within-community edges $f = 0.92$. This finding indicates that highly modular networks are most optimizable through cross-cluster weight tuning.

\begin{figure*}[t]
\begin{subfigure}[t]{0.015\textwidth}
  \vspace{-5cm}
  {\myfont \Large A}
\end{subfigure}
  \adjustbox{minipage=1em}{\label{sfig:testa}}%
  \begin{subfigure}[t]{\dimexpr.29\linewidth}
  \centering
  \vspace{-3.5cm}
      \includegraphics[trim = 30 0 0 50, scale=.45]{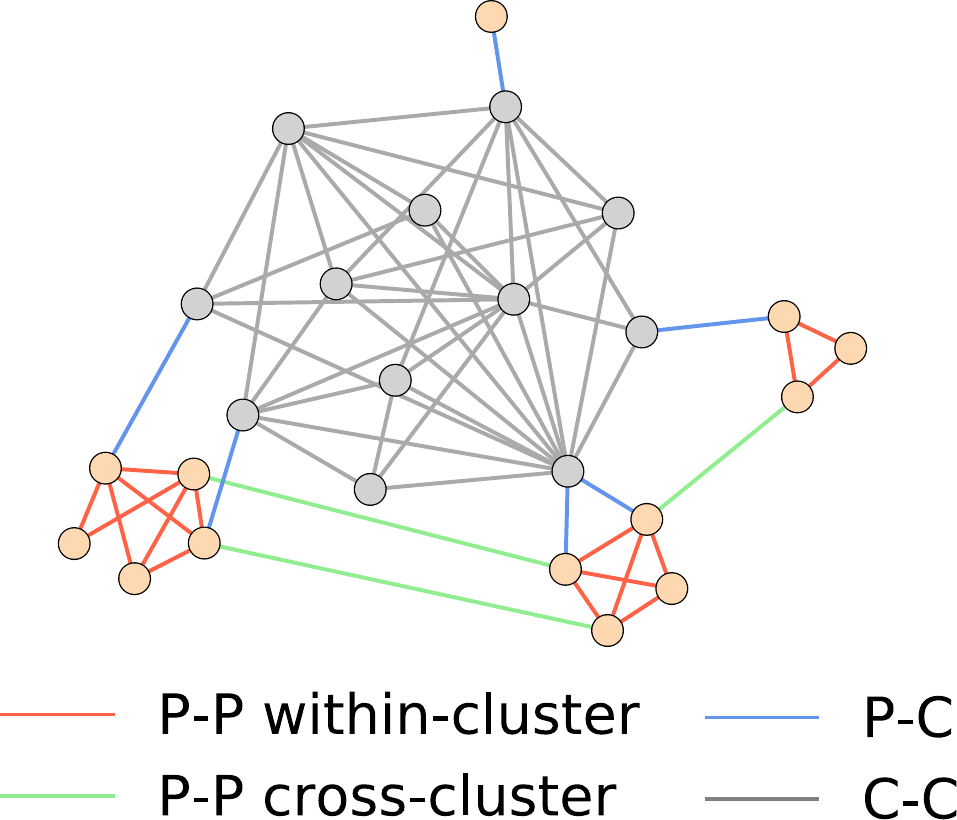}
  \end{subfigure}%
  \begin{subfigure}[t]{0.015\textwidth}
  \vspace{-5cm}
  {\myfont \Large C}
\end{subfigure}
  \adjustbox{minipage=1em}{\label{sfig:testb}}%
  \begin{subfigure}[t]{\dimexpr.34\linewidth}
  \centering
        \includegraphics[trim = 40 15 0 0, scale=.41]{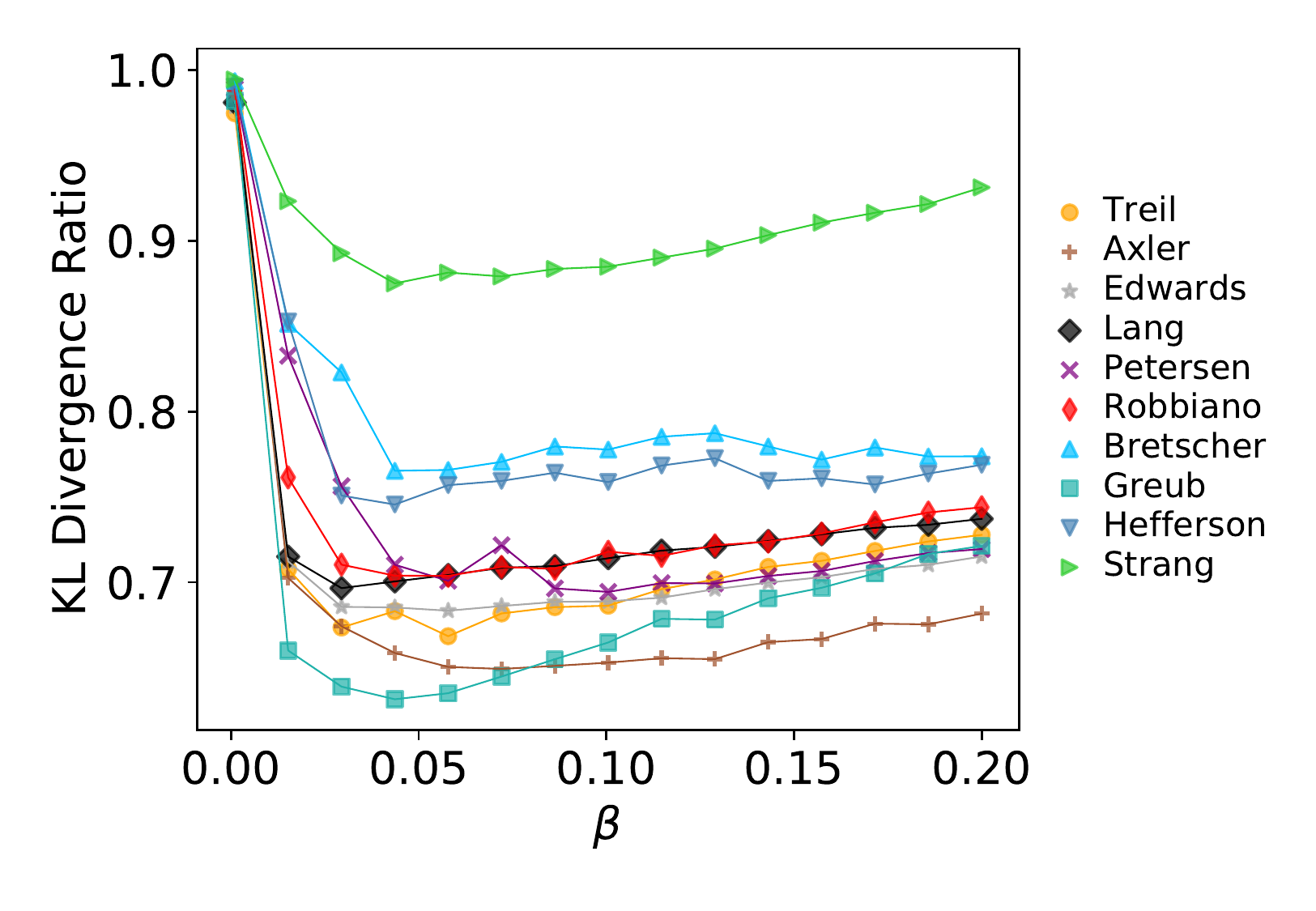}
  \end{subfigure}
  \begin{subfigure}[t]{0.015\textwidth}
  \vspace{-5cm}
  {\myfont \Large E}
\end{subfigure}
  \adjustbox{minipage=1em}{\label{sfig:testa}}%
  \begin{subfigure}[t]{\dimexpr.3\linewidth}
  \centering 
      \includegraphics[trim =30 15 0 0, scale=.40]{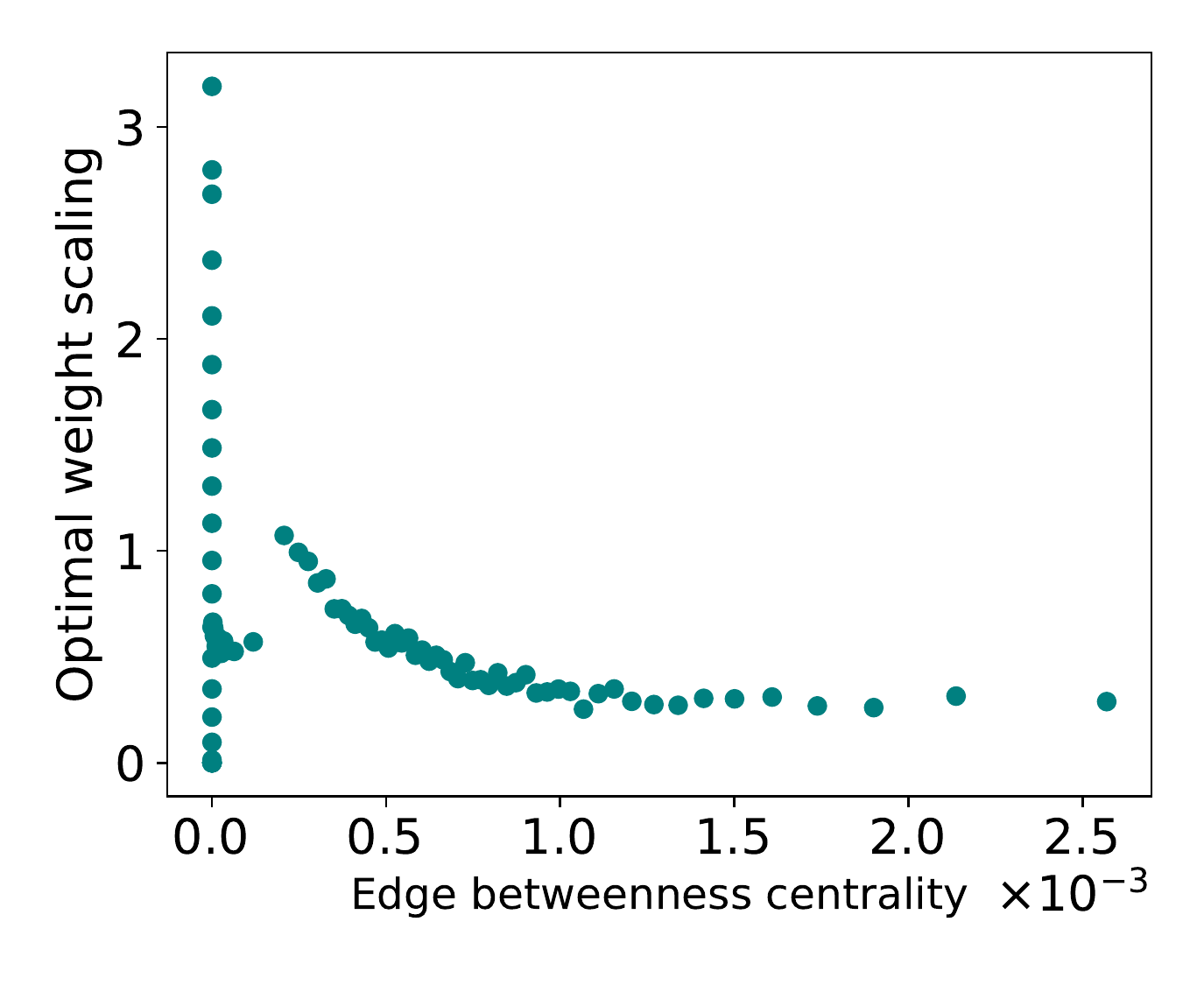}
  \end{subfigure}%
  \vspace{-1cm}
  \begin{subfigure}[t]{0.015\textwidth}
  \vspace{.5cm}
  {\myfont \Large B}
\end{subfigure}
  \adjustbox{minipage=1em}{\label{sfig:testa}}%
  \begin{subfigure}[t]{\dimexpr.29\linewidth}
  \vspace{2.8em}
  \centering
  \includegraphics[trim = 50 0 0 0, scale=.41]{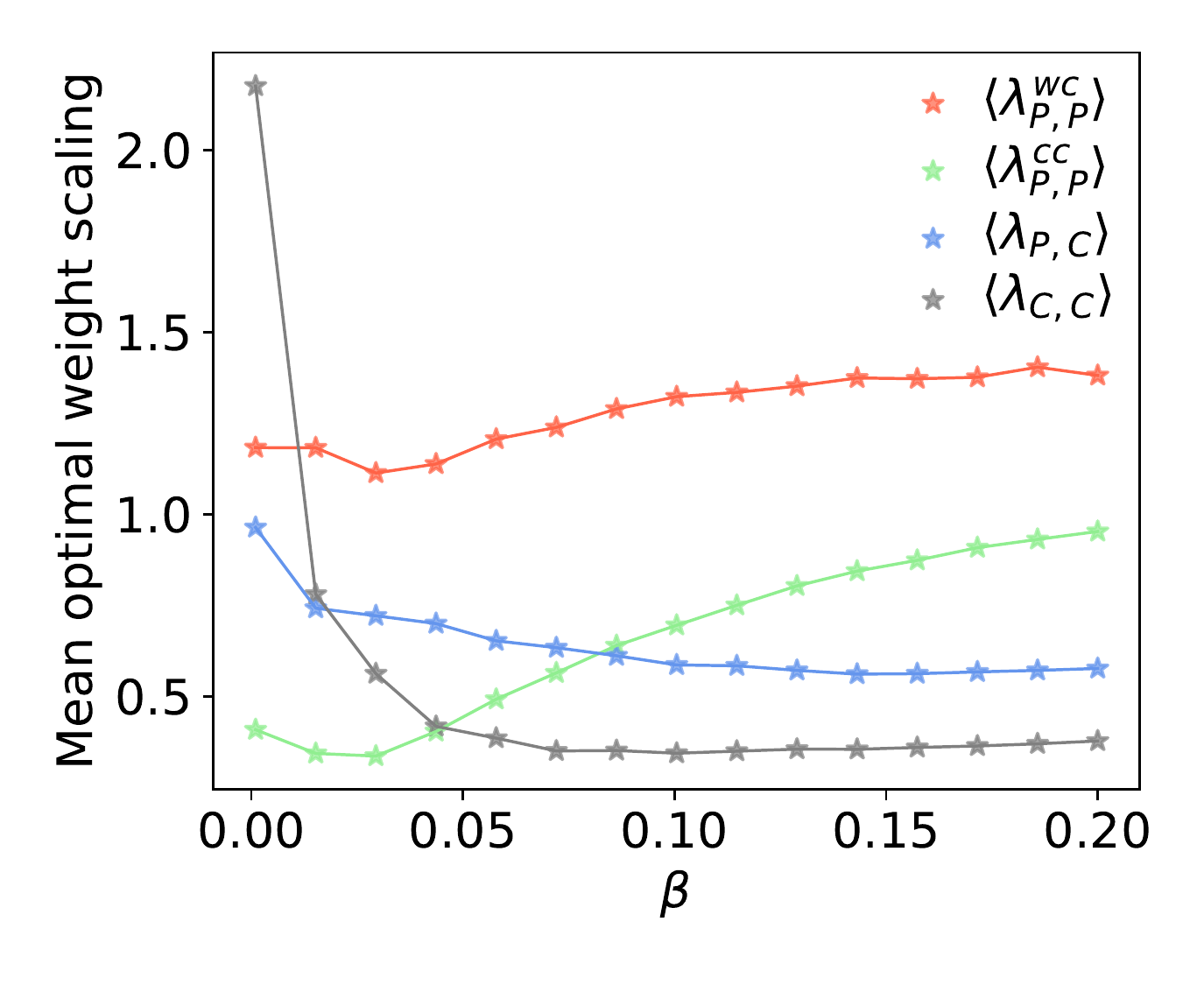}
  \end{subfigure}%
  \begin{subfigure}[t]{0.015\textwidth}
  \vspace{.5cm}
 {\myfont \Large D}
\end{subfigure}
\adjustbox{minipage=1em}{\label{sfig:testa}}%
  \begin{subfigure}[t]{\dimexpr.325\linewidth}
  \vspace{3.3em}
  \centering
      \includegraphics[trim = 30 40 0 0, scale=.39]{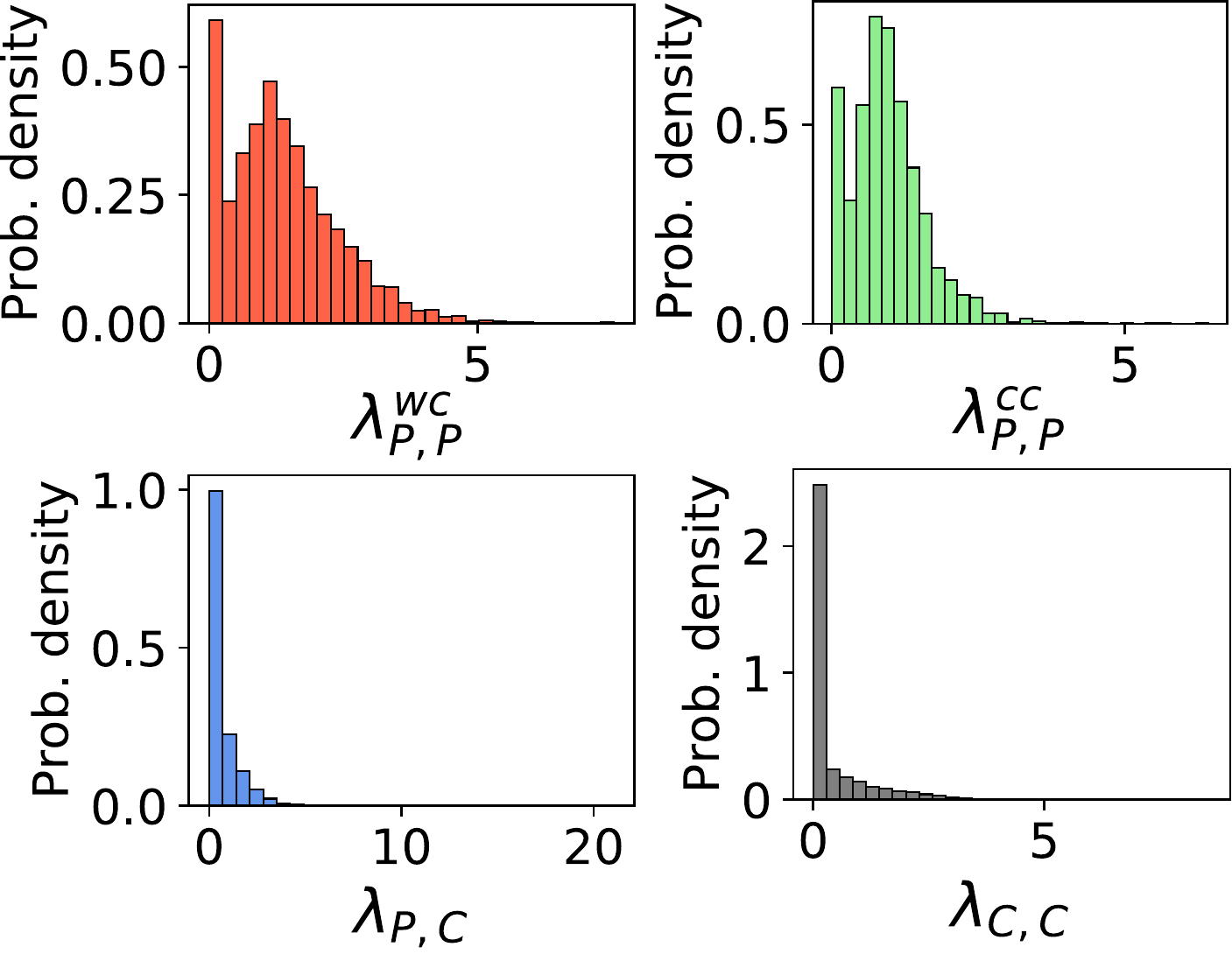}
  \end{subfigure}%
  \begin{subfigure}[t]{0.015\textwidth}
  \vspace{.5cm}
 {\myfont \Large F}
\end{subfigure}
\adjustbox{minipage=1em}{\label{sfig:testa}}%
  \begin{subfigure}[t]{\dimexpr.32\linewidth}
  \vspace{2.5em}
  \centering
      \includegraphics[trim = 30 0 0 0, scale=.423]{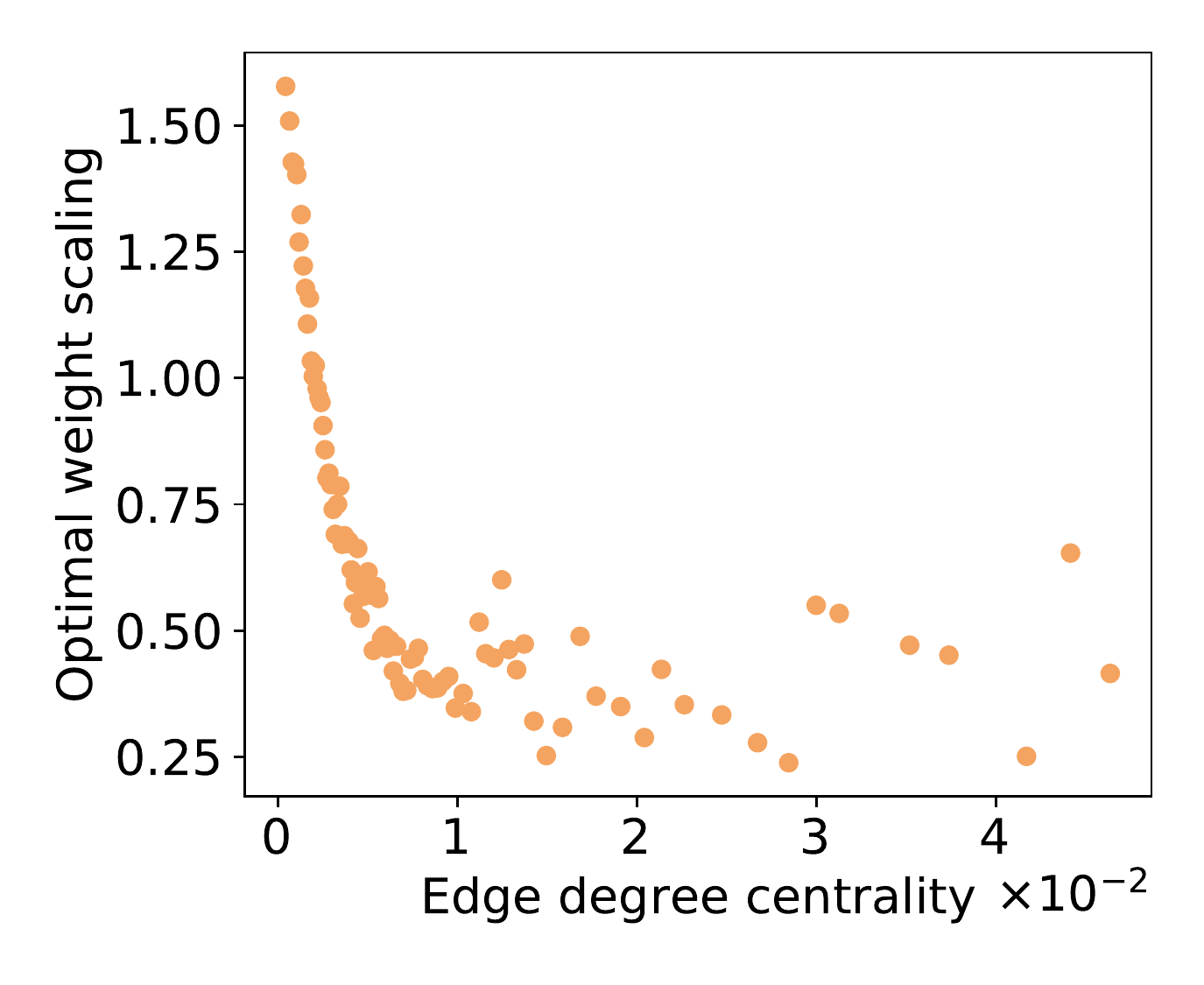}
  \end{subfigure}%
      \caption{\textbf{Optimizing the learnability of semantic networks extracted from college mathematics textbooks.} \emph{(A)} A schematic of how edges in the semantic networks were classified based on core-periphery node classification and periphery community structure. \emph{(B)} The optimal weight scaling for each of the four classes of edges shown at different values of $\beta$, averaged over all semantic networks. \emph{(C)} The Kullback-Leibler divergence ratio $\frac{D_{KL}(A||f(A_{\text{in}}))}{D_{KL}(A||f(A))}$ achieved with optimized weight scaling at different values of $\beta$. Results are shown for each of the ten semantic networks corresponding to the ten college-level linear algebra textbooks. \emph{(D)} The distribution of optimized edge weight scalings for the classes of edges at $\beta = 0.2$, aggregated over all semantic networks. \emph{(E,F)} The optimal edge weight scaling versus edge betweenness centrality and edge degree centrality, respectively, aggregated over all semantic networks for $\beta = 0.2$. Each datapoint represents an average over $500$ edges binned by centrality score. }
  \label{fig:4}
\end{figure*}

\paragraph{Watts-Strogatz networks.} Just as we generalized the analysis of modular networks to stochastic block networks, we can also extend our analysis of lattice networks to a wide range of randomly-generated Watts-Strogatz networks (SI Appendix, Fig. S3). Consistent with previous analysis, we find that edges in Watts-Strogatz networks that contribute to local, lattice-like clustering are emphasized when maximizing learnability.

\subsection*{Optimizing the learnability of semantic networks extracted from mathematics textbooks} Our results thus far have demonstrated that, for many classes of synthetic networks, edges that contribute to local clustering or intramodular connections are reinforced to maximize learnability. Still, it remains to be demonstrated that these results extend more generally to real-world information networks. To probe the optimal emphasis modulation strategies of real-world networks, we study semantic networks extracted from college-level linear algebra textbooks \cite{Christianson_Sizemore_Blevins_Bassett_2020}. Specifically, nodes represent recurring concepts (e.g., ``vector space'', ``invertible''), and edges between concepts are weighted by the number of sentences in which the two concepts co-occur. 

In previous analyses, we were able to reduce the number of free parameters in the network learnability optimization process by considering either network symmetry or a partitioning of the edges into classes that are made distinguishable by the network generation process (e.g., cross-cluster edges or non-ring edges). However, given that these semantic networks are empirical, when optimizing a network representation to maximize learnability, we cannot reduce the number of optimization parameters \emph{a priori}, and instead we must vary all edges with nonzero weight as free parameters. Specifically, for some semantic network $G_{SEM} = (V, E)$ with edge weights $w_{ij}$ and normalized transition matrix $A_{SEM}$, we determine a weighted graph $G_{\text{in}} = (V,E)$ with weights $w^{*}_{ij}$ such that its corresponding normalized transition matrix $A_{\text{in}}$ minimizes $D_{KL}(A_{SEM}||f(A_{\text{in}}))$. In particular, for every edge $e = \{i,j\} \in E$, the factor $\lambda _{ij} = \frac{w^{*}_{ij}}{w_{ij}}$ by which the edge $e$ is scaled in $G_{\text{in}}$ is used as an optimization parameter. Then, we reduce the number of free parameters by $1$ by enforcing the requirement that the total sum of edge weights in $G_{\text{in}}$ equals that of $G_{SEM}$. Doing so allows for more interpretable comparisons between the optimized network $A^*$ and the original semantic network $A_{SEM}$.

Interestingly, we find that for these semantic networks, very little improvement in learnability can be achieved at extremely low values of $\beta (\approx 0$, reflecting poor learning accuracy), but significant enhancement of learnability is possible for all other values of $\beta$ (Fig. \ref{fig:4}\emph{C}). This observation contrasts greatly with our prior findings in studying modular networks: that the greatest improvements in learnability occur near $\beta \approx 0$, with significantly diminishing improvements at higher $\beta$ values. One natural explanation for this difference is that the semantic networks do not possess Large-scale community structure, but rather can be characterized as possessing core-periphery structure with community structure within the periphery nodes \cite{Christianson_Sizemore_Blevins_Bassett_2020}. Therefore, the learnability optimization strategies for modular networks---which mainly involved over-emphasizing edges within clusters and de-emphasizing cross-cluster edges---are likely not applicable to increasing learnability of these semantic networks. 

To understand which edges are reinforced or de-emphasized to increase learnability, for each semantic network, we follow the procedure outlined in Ref.  \cite{Christianson_Sizemore_Blevins_Bassett_2020} and classify nodes into core and periphery categories (see Methods). Then, for each network, we determine the community structure of the periphery nodes, and categorize the edges of each network into four categories (Fig. \ref{fig:4}\emph{A}): edges between core nodes, edges between a core node and a periphery node, cross-cluster edges between periphery nodes, and within-cluster edges between periphery nodes. For each class of edges, we compute the mean optimal weight scaling over all ten semantic networks for $0 < \beta \leq 0.2$ (Fig. \ref{fig:4}\emph{B}). Notably, we find that for $\beta > 0.05$, edges between core nodes are de-emphasized the most, whereas edges between periphery nodes are reinforced. This finding is sensible as the core of a core-periphery network is densely connected. Therefore, any particular edge within the core could be de-emphasized to suppress potential spurious connections to nearby periphery nodes resulting from inaccurate learning. In addition, among the two classes of periphery-periphery edges, those that connect two nodes within the same periphery community tend to be exaggerated, whereas cross-cluster periphery-periphery edges are de-emphasized. These observations regarding cross-cluster and within-cluster periphery-periphery edges are consistent with our prior analyses of optimal cross-cluster edge weights in modular networks, which also suggest that cross-cluster weights should be de-emphasized.

We can further characterize the types of edges that are either over- or under-emphasized by comparing changes in edge weight with structural measures of centrality (see Methods). First, we consider the relationship between edge weight scaling and edge betweenness centrality, a metric that quantifies the frequency with which shortest paths pass through a given edge. Given that cross-cluster edges have high edge-betweenness, it is natural to expect that edges with high betweenness will be de-emphasized when optimizing learnability. Indeed, we observe that, aside from edges with an edge betweenness centrality of $0$, there is a clear inverse relationship between edge betweenness centrality and optimal edge weight scaling in semantic networks (Figs \ref{fig:4}\emph{E}; $\beta = 0.2$). 

We also consider edge weight scaling and edge degree centrality, a metric that quantifies the average weighted degree of the two connected nodes. Edges within the core of a core-periphery network are likely to be incident on nodes with greater connectivity, and thus would generally have higher edge degree centrality. Thus, we expect that edge degree centrality will be inversely related to optimal edge weight scaling. Consistent with these expectations, we observe that edges with lower edge degree centrality tended to be reinforced more, and edges with higher edge degree centrality tended to be de-emphasized more (Fig. \ref{fig:4} \emph{F}). 

\begin{figure*}[t]
\vspace{1.5em}
\begin{subfigure}[t]{0.015\textwidth}
  \vspace{-5.3cm}
  {\myfont \Large A}
\end{subfigure}
  \adjustbox{minipage=1em}{\label{sfig:testa}}%
  \begin{subfigure}[t]{\dimexpr.31\linewidth}
  \centering
      \includegraphics[trim = 30 0 0 0, scale=.42]{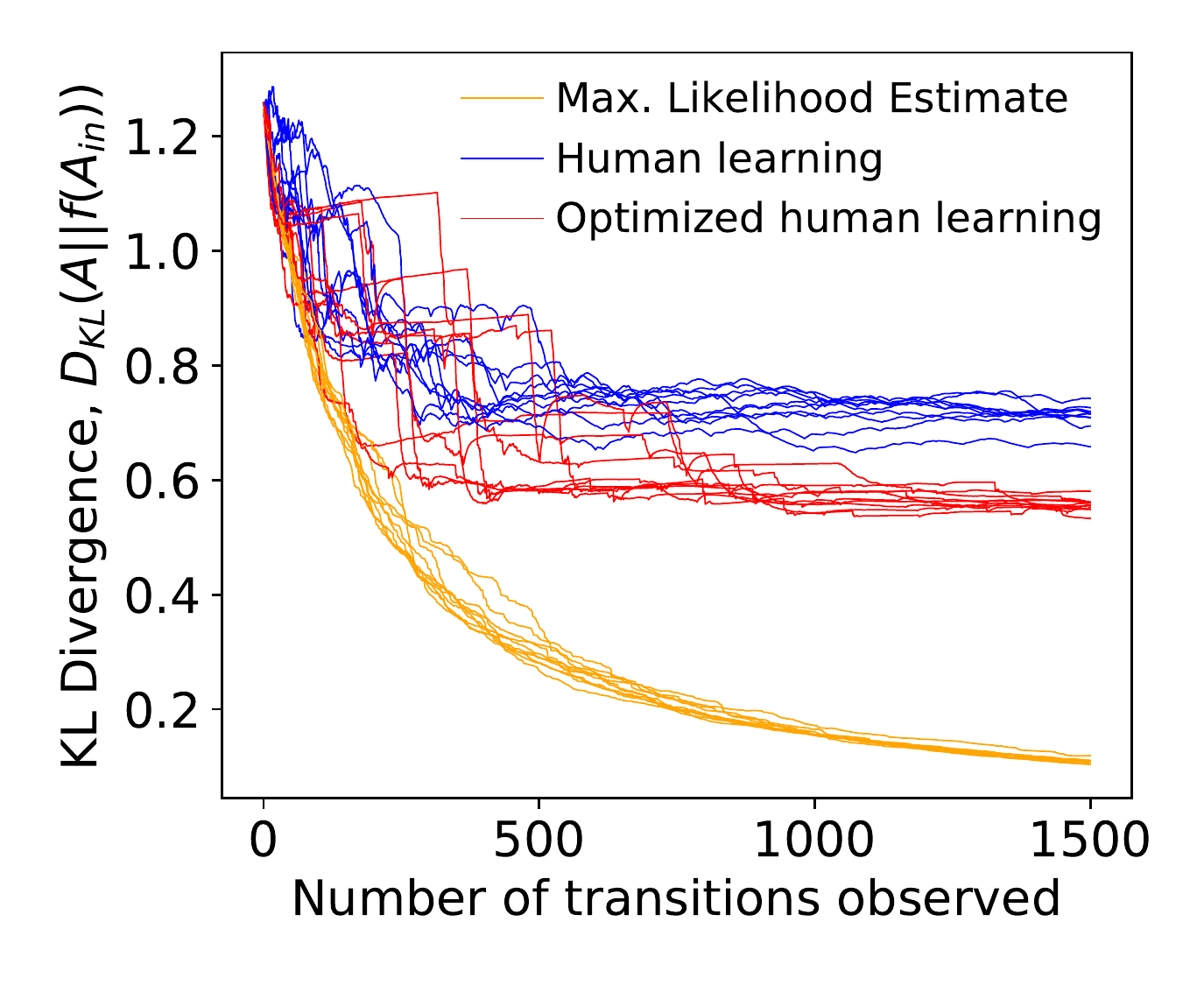}
  \end{subfigure}%
  \begin{subfigure}[t]{0.015\textwidth}
  \vspace{-5.3cm}
  {\myfont \Large B}
\end{subfigure}
  \adjustbox{minipage=1em}{\label{sfig:testb}}%
  \begin{subfigure}[t]{\dimexpr.31\linewidth}
  \centering
        \includegraphics[trim = 30 0 0 0, scale=.42]{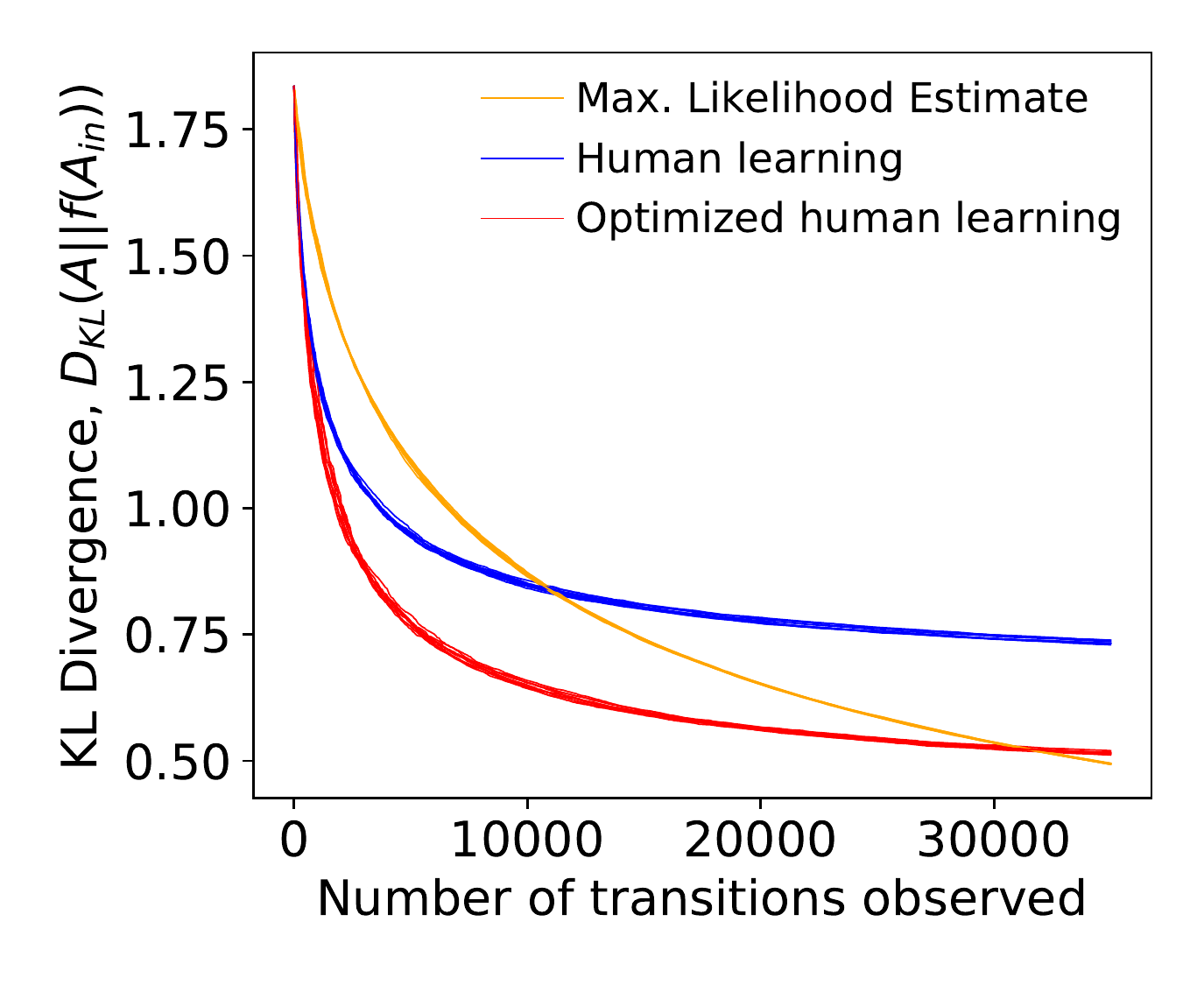}
  \end{subfigure}
  \begin{subfigure}[t]{0.015\textwidth}
  \vspace{-5.3cm}
  {\myfont \Large C}
\end{subfigure}
  \adjustbox{minipage=1em}{\label{sfig:testa}}%
  \begin{subfigure}[t]{\dimexpr.31\linewidth}
  \centering 
      \includegraphics[trim =30 0 0 0, scale=.42]{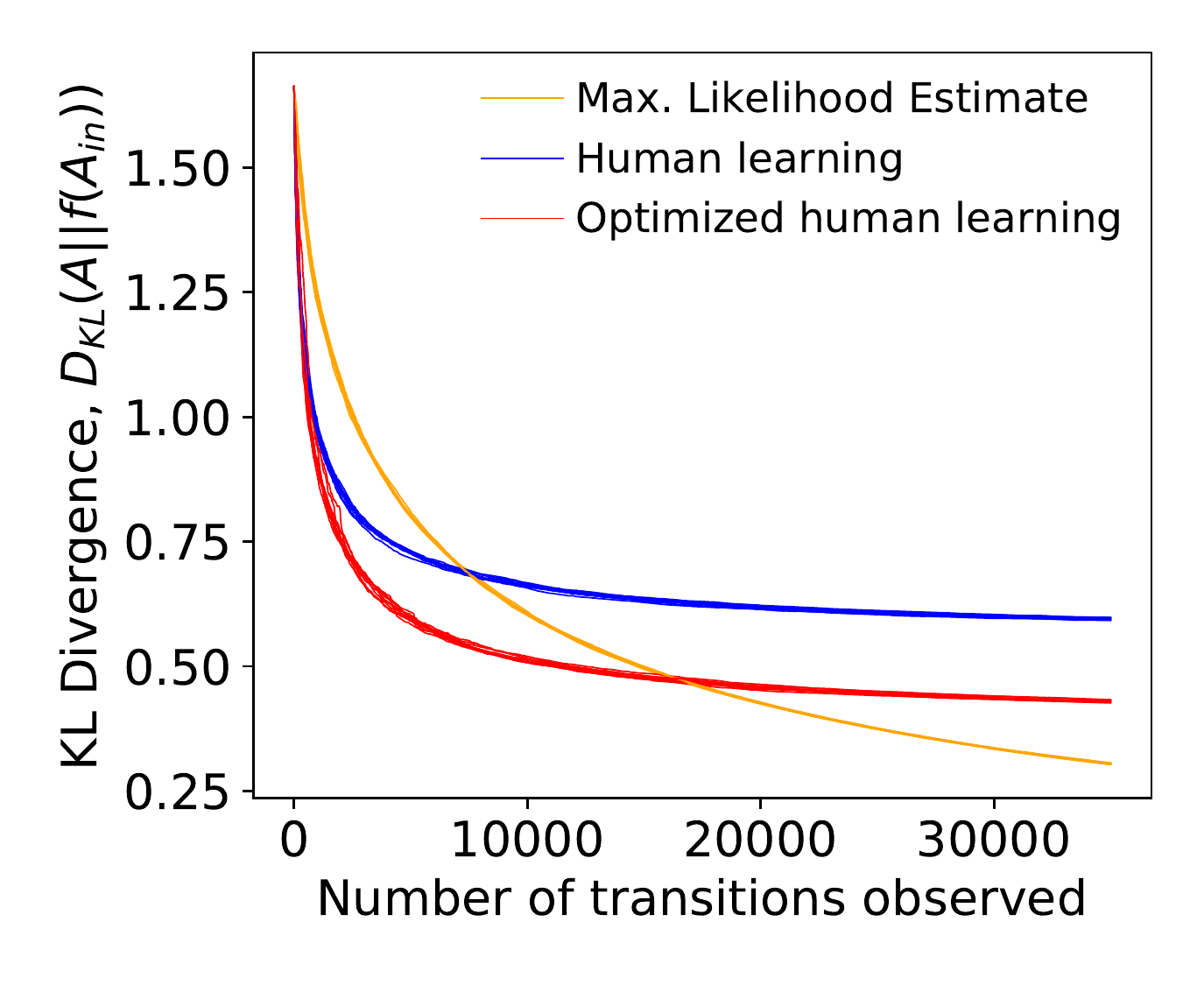}
  \end{subfigure}%
      \caption{\textbf{Performance of network learning strategies in the transient regime.} Each panel shows the Kullback-Leibler divergence between some true network and the learned network as a function of the number of observed transitions. Three network learning strategies are shown: maximum likelihood estimation (optimal in the limit of infinite observations), standard human network learning (supported by Ref. \cite{Lynn_Kahn_Nyema_Bassett_2020}), and optimized human network learning (introduced in this paper). All plots report ten simulations of each network learning strategy, with human learning and optimized human learning simulations run at $\beta = 0.1$, close to the median learning accuracy reported in Ref. \cite{Lynn_Kahn_Nyema_Bassett_2020}. The three networks analyzed are: \emph{(A)} the modular network with $15$ nodes (see Fig.\ref{fig:1}\emph{A}), \emph{(B)} the semantic network extracted from the linear algebra textbook authored by Axler, and \emph{(C)} the semantic network extracted from the linear algebra textbook authored by Edwards.
     }
  \label{fig:5}
\end{figure*}

\subsection*{Performance of network optimization strategies in the transient network learning regime} Thus far, our analysis has implicitly assumed that human network learners are allowed to observe infinite sequences of network transitions \cite{Lynn_Kahn_Nyema_Bassett_2020,Lynn_Papadopoulos_Kahn_Bassett_2020}. While this assumption greatly simplifies optimization strategies, it is possible that such strategies break down when the number of transitions that human learners are allowed to observe is limited. To address this possibility, we ran simulations of the network learning process in the transient regime with three different network learning strategies (see Methods): 1) maximum likelihood estimation (optimal in the infinite observation limit), 2) standard human network learning (as reported in Ref. \cite{Lynn_Kahn_Nyema_Bassett_2020}), and 3) optimized human learning (as described here in our paper). Both the standard and optimized human learning strategies were evaluated at $\beta = 0.1$, which is close to the mean learning accuracy reported in Ref. \cite{Lynn_Kahn_Nyema_Bassett_2020}. These simulations were run for the modular graph with $15$ nodes (Fig. \ref{fig:1}\emph{A}) and the semantic networks extracted from the linear algebra textbooks authored by Axler and Edwards (Fig. \ref{fig:1}\emph{B} and Fig. \ref{fig:1}\emph{C}, respectively).

For the modular network, optimized human learning maintained an edge in accuracy over standard human network learning as the number of transitions observed increased (Fig. \ref{fig:5}\emph{A}). Both learning strategies were outperformed by maximum likelihood estimation throughout the duration of the simulated learning processes. Remarkably, for the semantic networks, both human learning and optimized human learning initially outperformed the accuracy of maximum likelihood estimation, with optimized human learning maintaining its superiority for a longer duration (Fig. \ref{fig:5}\emph{B,C}). One plausible explanation is that the inductive biases introduced in the human learning process enable humans to initially learn clustered areas of networks more efficiently than unbiased maximum likelihood estimation strategies. 

\section*{Discussion}

In this article, we study how networks presented to human learners can be tuned to increase learnability. In particular, we compute the optimal network to present to a human learner so as to minimize the discrepancy between the learned representation and the target network. First, these methods were used to analyze two simple networks: a modular graph and a lattice graph. We find that for both graphs, improvements in learnability can be made by de-emphasizing edges that connect different modules or clusters of nodes, and by exaggerating edges within modules or small clusters. This finding is consistent with studies of \emph{in silico} models and \emph{in vivo} animal behavior of sampling spaces. Animals and computational models exhibit non-random patterns of exploration in order to better sample an environment with non-regular network structure \cite{McNamee2021}, effectively emphasizing and over-representing specific harder-to-learn portions of the environment \cite{Whittington2020}. Further, these improvements increase considerably in magnitude for highly inaccurate human learners, but are less advantageous for accurate learners. Importantly, for inaccurate learners, the optimal input networks for both modular and lattice graphs result in internal network representations that capture clusters in the original network in a near perfect manner, but poorly capture edges between small clusters or modules. Notably, edges between communities or clusters are already naturally difficult to learn in the absence of disproportionate edge weighting in the input network \cite{Lynn_Kahn_Nyema_Bassett_2020, Lynn_Papadopoulos_Kahn_Bassett_2020}, but are found to be worth de-emphasizing further when optimizing overall learnability. Our findings are consistent with prior work showing that the difficulty of learning cross-cluster edges in modular networks is robust to the size and number of modules in the network \cite{Karuza_Kahn_Bassett_2019}.

Then, to probe whether our findings with the modular and lattice networks extended more generally to Larger, more complex networks lacking a high degree of structural symmetry or uniform transition probabilities, we analyzed the optimization of learnability of networks generated from generative network models. We first began by studying the optimization of stochastic block networks. Importantly, we observe that for stochastic block networks with a high fraction of edges within communities, significant gains in network learnability can be achieved only by tuning a single parameter representing the weight of all cross-cluster edges. Specifically, we observe that stochastic block networks optimized for learnability de-emphasized cross-cluster edges. Next, motivated by the prevalence of heterogeneous degree distributions among real-world networks \cite{Lerman_Yan_Wu_2016, Barabasi_Albert_1999, Nekovee_Moreno_Bianconi_Marsili_2007,albert_2005, tadic_mitrovic_2009}, we investigate the optimization of learning for degree-corrected stochastic block networks. Through applying a similar single-parameter optimization approach, we find that degree-corrected stochastic block networks share similar learning optimization properties to standard stochastic block networks when tuning cross-cluster edge weights. However, the efficacy of these optimization strategies in improving network learnability was found to be slightly higher for degree-corrected stochastic block networks. This finding suggests that the learning of networks with hierarchically modular organization can be improved significantly more (using this cross-cluster edge weight tuning) than can random modular networks. Taken together with prior work showing that hierarchically modular networks share similar information-theoretic properties with a Large class of real-world networks \cite{Lynn_Papadopoulos_Kahn_Bassett_2020} that random modular networks lack, these findings have implications for how features of real-world information networks ought to be weighted or designed.

Finally, to understand optimal strategies for enhancing the learnability of real-world information networks, we analyze how semantic networks extracted from college-level mathematics textbooks can be re-weighted to maximize learnability. These networks exhibit core-periphery structure, indicating that they are composed of nodes that can roughly be divided into a densely-connected core and a periphery that is loosely connected to nodes within the core \cite{Csermely_London_Wu_Uzzi_2013, Verma_Russmann_Araujo_Nagler_Herrmann_2016, Rombach_Porter_Fowler_Mucha_2014}. In addition, prior work has established that the periphery of these semantic networks possesses community structure \cite{Christianson_Sizemore_Blevins_Bassett_2020}. Importantly, unlike modular networks, we find that the semantic networks are not very optimizable near $\beta \approx 0$, but are significantly optimizable even for moderately Large values of $\beta$.
One explanation for the difference in optimization near $\beta = 0$ between the modular networks and semantic networks is that, for modular networks, optimal input networks approach disconnected graphs as $\beta \rightarrow 0$. Typically, as $\beta \rightarrow 0$, the learned representation from any input network approaches a uniform network with no particular structure \cite{Lynn_Kahn_Nyema_Bassett_2020,Lynn_Papadopoulos_Kahn_Bassett_2020}. However, a competing limit may occur when studying the optimization of learnability of modular networks, in that cross-cluster weights may also approach $0$. In this limit, a network presentation approaches disconnected components, with each component representing a module of the original network. Thus, when $\beta$ is taken close to $0$, learning inaccuracies primarily strengthen edges within modules, causing only minor decreases in overall accuracy of the optimized network as nodes within modules or clusters are already densely connected with each other in the original target network. In contrast, for networks without overall community structure such as the semantic networks extracted from the mathematical texts, the optimal input graph always remains connected as $\beta \rightarrow 0$, and thus any learned representation will approach an all-to-all network with uniform transition probabilities in this limit.

To characterize the features of these semantic networks that become reinforced or de-emphasized when optimizing for learnability, we categorize the edges in each of the semantic networks by the core-periphery status of their endpoints, as well as by cross-cluster or within-cluster participation in periphery community structure. We find that edges contained solely amongst the periphery of these networks are emphasized the most, whereas core edges are de-emphasized, even at higher $\beta$. These findings suggest that the learnability of information networks can generally be enhanced by placing additional emphasis on relationships between less commonly occurring concepts, and de-emphasizing highly central concepts. To assess this idea further, we investigated how the centrality measures of edge betweenness and edge degree was associated with the scaling of edge weights for these semantic networks. In particular, we find that both centrality measures are negatively correlated with optimal edge weight scaling, confirming that in these semantic networks, edges that are highly central generally should be de-emphasized to maximize learnability.

~\\ \noindent \textbf{Methodological Considerations:} We note that our results on optimizing the learnability of generated networks are strictly a lower bound on improvements in human learnability of networks that can be afforded through targeted emphasis and de-emphasis of particular input network features. Since we only consider optimization of learnability of generated networks via single-parameter tuning (cross-cluster edge weights or non-ring edge weights), it is likely that more nuanced emphasis modulation strategies may enhance learnability even further beyond what has been demonstrated for both classes of stochastic block networks analyzed, as well as for Watts-Strogatz networks. Similarly, our results on semantic networks extracted from mathematics textbooks only represent lower bounds on improvements in learnability that can be achieved through targeted emphasis modulation. This is a consequence of the fact that each of the semantic networks analyzed had thousands of edge weights to be varied as parameters, and thus achieving globally optimal network representations via dual annealing was not always possible. In addition, the analyses presented in this work have yet to consider whether adding edges that are entirely nonexistent in a target network to the input network presented to a human learner may enhance learnability of the target network. Preliminary findings suggest that increases in learnability can indeed occur when nonexistent edges are added to an input network representation (SI Appendix, Fig. S4). Future work could fruitfully consider how relaxing edge-existence constraints may affect the efficacy of the estimated optimal network representations. 

\section*{Conclusion}
Recent advances in the study of human information processing have shed light on the ways that humans learn information networks. Rather than mapping the structure of networks exactly, inaccuracies in human learning often result in erroneous or biased internal representations. To overcome inaccuracies in human learning, we investigate how networks presented to human learners can be designed to counteract and minimize inaccuracies in learning. Across a range of synthetic networks, we find that reinforcing edges within clusters and de-emphasizing edges between clusters improves network learnability. In addition, we analyze how real-world semantic networks can be optimized for learnability, thereby uncovering the fact that relationships between periphery concepts ought to be reinforced. Together, our findings demonstrate that the learnability of network representations can be significantly enhanced through intentionally modulating the emphasis of specific network features.

\section*{Materials and Methods}
\subsection*{Converting between weighted graphs and transition matrices}
To convert between graphs and transition matrices, we use the following two relations: 

\noindent Given a weighted graph $G$, the normalized transition matrix corresponding to random walks is given by $A_{ij} = \frac{G_{ij}}{\sum_{k = 1}^{N} G_{ik}}$. 

\noindent Conversely, the adjacency matrix for a weighted, undirected graph corresponding to a reversible transition matrix $A$ is determined up to a multiplicative constant by $G_{ij} = \pi_{i}A_{ij}$, where $\pi$ is the stationary distribution of $A$. 

\subsection*{Optimization methods}

To assess how much a learned network differs from the target network, we use the Kullback-Leibler divergence. The Kullback-Leibler divergence between normalized transition networks $A$ and $B$ is defined as 
\begin{equation}
    D_{KL}(A||B) = - \sum_{i}\pi _i \sum_{j} A_{ij} \log \bigg(\frac{B_{ij}}{A_{ij}}\bigg),
\end{equation}
where $\pi$ is the stationary distribution of $A$, and only terms with non-zero transition probabilities are summed over. 

For some target transition structure $A$, the optimal input structure $A_{\text{in}} = A^{*}$ was determined using the dual annealing optimization method in $scipy$, with $D_{KL}(A||f(A_{\text{in}}))$ as the cost function. During the optimization process, the edge weights of $A_{\text{in}}$ corresponding to edges that exist in $A$ were varied as free parameters bounded between $0$ and $1$. The matrix $A_{\text{in}}$ was then normalized prior to every cost function evaluation. For the modular and lattice graphs, the number of free parameters varied during optimization was reduced based on the networks' respective symmetries. Network symmetries were computed using the $iGraph$ package in Python \cite{igraph}.  

\subsection*{Network generation}
We generated stochastic block networks as follows: Starting with $N$ nodes, we first assign each node to a community labelled $1$ through $5$ uniformly at random. Then, $\lfloor|E|p \rfloor$ edges are added between nodes in the same module, where $|E|$ is the total number of edges in the network. Specifically, each addition of a within-cluster edge was performed by selecting a community uniformly at random, selecting two non-adjacent nodes within the community uniformly at random, and then adding an edge between the selected nodes. Finally, $|E| - \lfloor|E|p \rfloor$ edges are then added between nodes in different communities. This process was performed by selecting two different communities uniformly at random, and then selecting one node in each community uniformly at random, and connecting the nodes if they are non-adjacent. If they are adjacent, the step is repeated. 

For degree-corrected stochastic block networks, a similar network generation method was used, with modifications adapted from the procedure described in Ref.  \cite{Lynn_Papadopoulos_Kahn_Bassett_2020}. Specifically, each node is assigned an index from $i$ from $1$ to $200$, and is then assigned a weight $i^{-1}$. Then, the probability of a given node being chosen for the addition of an edge at any step was proportional to the weight of the node. 

\subsection*{Analysis of semantic networks}
Given a weighted network $G = (V, E)$ with edge weights $w_{ij}$, we identify core-periphery structure by finding a partition of the vertex set $V$ into disjoint sets $C$ and $P$ so that the following core-ness quality function is maximized:
\begin{equation}
    Q_C = \frac{1}{v_C}\left(\sum_{i, j \in C}(w_{ij} - \gamma _C \overline{w}) - \sum _{i, j \in P} (w_{ij} - \gamma _C \overline{w})\right),
\end{equation}
where $v_C$ is a normalization constant, $\overline{w}$ is the average over all edge weights, and $\gamma _C$ is a resolution parameter that we set to $1$. 

\noindent Given a partition $V = C \cup P$ into core and periphery nodes, we evaluate the community structure of the periphery of $G$ by maximizing the following modularity quality function on the subgraph of $G$ induced on $P$:
\begin{equation}
    Q_C = \frac{1}{v_M}\left(\sum_{i, j \in P}(w_{ij} - \gamma _M \frac{s_i s_j}{v_M})\delta _{ij}\right),
\end{equation}
where $v_M$ is a normalization constant, $s_i = \sum _{j \in P} w_{ij}$ is the sum over the weights of all edges incident on vertex $i$, and $\gamma _M$ is a resolution parameter that we set to $1$.

\noindent The edge betweenness centrality for some edge $e$ is defined as
\begin{equation}
    C_B(e) = \frac{2}{(n-1)(n-2)}\sum_{s,t \in V} \frac{\sigma (s,t |e)}{\sigma (s, t)},
\end{equation}
where $\sigma(s,t)$ is the number of weighted shortest paths between nodes $s$ and $t$, and $\sigma (s,t|e)$ is the number of such shortest paths that go through the edge $e$. 

\noindent For an edge $e = \{i, j\}$, we define the weighted edge degree centrality as 
\begin{equation}
    C_D(e) = \frac{s_i + s_j}{2 \sum_{p < q} w_{pq}},
\end{equation}
where $s_i = \sum_{k} w_{ik}$ is the weighted degree of node $i$ in the network. 

\subsection*{Simulating transient network learning}
While observing random walks drawn from some transition network, the maximum likelihood estimate for the transition network can be described by 
\begin{equation}
    A^{\text{MLE}}_{ij}(t) = \frac{n_{ij}(t)}{\sum_k n_{ik}(t)} ,
\end{equation}
where $n_{ij}$ is the number of observed transitions from node $i$ to node $j$ by time step $t$.
As described in Ref. \cite{Lynn_Kahn_Nyema_Bassett_2020}, the human learned representation takes a similar form:
\begin{equation}
    \Tilde{A}_{ij}(t) = \frac{\Tilde{n}_{ij}(t)}{\sum_k \Tilde{n}_{ik}(t)}, 
\end{equation}
where $\Tilde{n}_{ij}$ is the revised (and erroneous) count of transitions from node $i$ to node $j$. 
In particular, it is described by 
\begin{equation}
    \Tilde{n}_{ij}(t+1) = \Tilde{n}_{ij}(t) + B_t(i)[j=x_{t+1}],
\end{equation}
where $x_{t+1}$ is the node observed at time step $t+1$, $[\cdot]$ is the Iverson bracket, and $B_t(i)$ encodes an internal belief about which node was observed at time $t$, described by 
\begin{equation}
B_t(i) = \frac{1}{Z}\sum_{\Delta t=0}^{t-1}e^{-\beta \Delta t}[i = x_{t-\Delta t}],
\end{equation}
where $Z$ represents a normalizing constant.

\section*{Acknowledgements}
The authors thank Christopher Kroninger for feedback on earlier versions of this manuscript. The authors also thank Pixel Xia for useful discussions on Sierpiński graph construction. This work was supported by the Army Research Office (DCIST-W911NF-17-2-0181) and the National Institute of Mental Health (1-R21-MH-124121-01). D.S.B. acknowledges additional support from the John D. and Catherine T. MacArthur Foundation, the ISI Foundation, an NSF CAREER Award PHY-1554488, and the Center for Curiosity. The content is solely the responsibility of the authors and does not necessarily represent the official views of any of the funding agencies.

\section*{Citation Diversity Statement} 
Recent work in several fields of science has identified a bias in citation practices such that papers from women and other minorities are under-cited relative to the number of such papers in the field \cite{Dworkin2020, maliniak2013gender, caplar2017quantitative, chakravartty2018communicationsowhite, YannikThiemKrisF.SealeyAmyE.FerrerAdrielM.Trott2018, dion2018gendered}. Here we sought to proactively consider choosing references that reflect the diversity of the field in thought, form of contribution, gender, and other factors. We obtained predicted gender of the first and last author of each reference by using databases that store the probability of a name being carried by a woman \cite{Dworkin2020,cleanbib}. By this measure (and excluding self-citations to the first and last authors of our current paper), our references contain 21.05\% woman(first)/woman(last), 7.89\% man/woman, 14.17\% woman/man, and 56.88\% man/man. This method is limited in that a) names, pronouns, and social media profiles used to construct the databases may not, in every case, be indicative of gender identity and b) it cannot account for intersex, non-binary, or transgender people. We look forward to future work that could help us to better understand how to support equitable practices in science.

\bibliography{main}
\bibliographystyle{unsrt}

\end{document}


\maketitle


\subsection*{Supplementary Results}
\paragraph{The lattice graph exemplar.} To understand how optimizing network learnability varies with the topology of the target network, we also consider the lattice graph presented in Fig.\ref{fig:2}\emph{A}. Unlike the modular graph, the lattice graph has only two structurally unique edges: edges within triangles and edges between triangles. Thus, an input network $A_{\text{in}}$ can be fully described by only one free parameter, the weight $\lambda_l$ of edges between triangles (orange), relative to the weight of edges within triangles (grey). Due to the reduction in the number of parameters, we are able to characterize gains in learnability in the lattice network while continuously varying both $\lambda_{l}$ and $\beta$ (Fig. \ref{fig:2}\emph{B}). Notably, learnability of the lattice graph increases markedly in the regime of low $\beta$ and low $\lambda_{l}$. Moreover, across all values of $\beta$, we find that optimizing learnability requires de-emphasizing the edges between triangles (Fig. \ref{fig:2}\emph{C}).

We note that there are considerable differences between the optimal emphasis modulation strategies of modular and lattice graphs as a function of $\beta$. First, the profiles of the curves of optimal edge weight values are significantly different between the two networks (compare main text Fig. 1\emph{E} and Fig. \ref{fig:2}\emph{C}). Specifically, the optimal edge weight curve of the lattice network shows an inflection point, whereas both optimal edge weight curves for the modular graph do not. Similar qualitative differences also appear between the optimal Kullback-Leibler divergence curves (compare main text Fig. 1\emph{F} and Fig. \ref{fig:2}\emph{D}). These differences arise despite the fact that both networks were chosen to share the same local properties (all nodes have $4$ neighbors), and thus have corresponding transition matrices with the same stationary distribution. This observation suggests that different networks require different approaches to maximize learnability, and that the efficacy of these approaches will differ by topology.

\paragraph{A Sierpiński graph exemplar.} To assess whether the strategy of over-emphasizing edges within clusters and de-emphasizing those between clusters extends to Larger networks with more complex community organization, we consider a modified version of the Sierpiński network with $3$ hierarchical levels and $5$ communities at each level. Specifically, the network was modified to include a sixth community at the highest level (Fig. \ref{fig:4}\emph{A}), allowing the graph to become $5$-regular, and therefore allowing its transition network to be uniform. This network was chosen to assess how edges at various levels ought to be weighted to maximize learnability in networks with hierarchically modular organization. Despite containing $150$ nodes, the network possesses a Large degree of structural symmetry, and has only four unique edges: level-2 cross-cluster edges ($\lambda _{cc}^2$, orange), level-3 cross cluster edges ($\lambda _{cc}^3$, blue), boundary edges adjacent to level-2 cross cluster edges ($\lambda _{b}^2$, green), and boundary edges adjacent to level-3 cross-cluster edges ($\lambda _{b}^3$, grey). As before, we reduce the number of free parameters by $1$ and fix $\lambda _{b}^3 = 1$.

Overall, we find that de-emphasizing both classes of cross-cluster edge weights is an effective strategy for optimizing the learnability of the Sierpiński network (Fig. \ref{fig:2}\emph{B,C}). However, there are slight differences in optimal edge weights between the level-2 and level-3 edges (Fig. \ref{fig:2}\emph{D}). In particular, we find that edges at the highest level of organization (level-3 edges) ought to be de-emphasized slightly more than level-2 edges. The efficacy of these optimization strategies scales similarly with $\beta$ as in the case of the $15$-node modular graph (compare Fig. \ref{fig:2}\emph{F} and Fig. \ref{fig:2}\emph{F}). The learned representations of the Sierpiński network with and without edge weight optimization are shown for $\beta = 0.05$ in Figs. \ref{fig:2}\emph{C} and \ref{fig:2}\emph{E}, respectively. These findings further suggest that the learnability of both modular and hierarchically modular networks can be substantially enhanced through the de-emphasis of cross-cluster edges, and the reinforcement of within-cluster edges. Moreover, by optimizing learnability, the learned representation of the hierarchically modular network maintains the fine-scale community structure (Fig. \ref{fig:2}\emph{E}). Interestingly, at low $\beta$ values, the learned representation resulting from optimal edge weights strikes a trade-off between local and global features: it strongly captures the features of each of the small $5$-node cliques, but poorly captures the hierarchical structure of the network. This pattern is likely a consequence of the fact that, at low $\beta$ values, near-perfect learning is impossible, and thus an optimal weighting strategy for minimizing the Kullback-Leibler divergence would place emphasis on accurately learning the most commonly occurring substructure.

\begin{figure}
\centering
\begin{subfigure}[t]{0.015\textwidth}
  \vspace{-5cm}
  {\myfont \Large A}
\end{subfigure}
  \adjustbox{minipage=1em}{\label{sfig:testa}}%
  \begin{subfigure}[t]{\dimexpr.34\linewidth}
  \centering
  \vspace{-5.3cm}
  \begin{overpic}[width=\textwidth,tics=10]{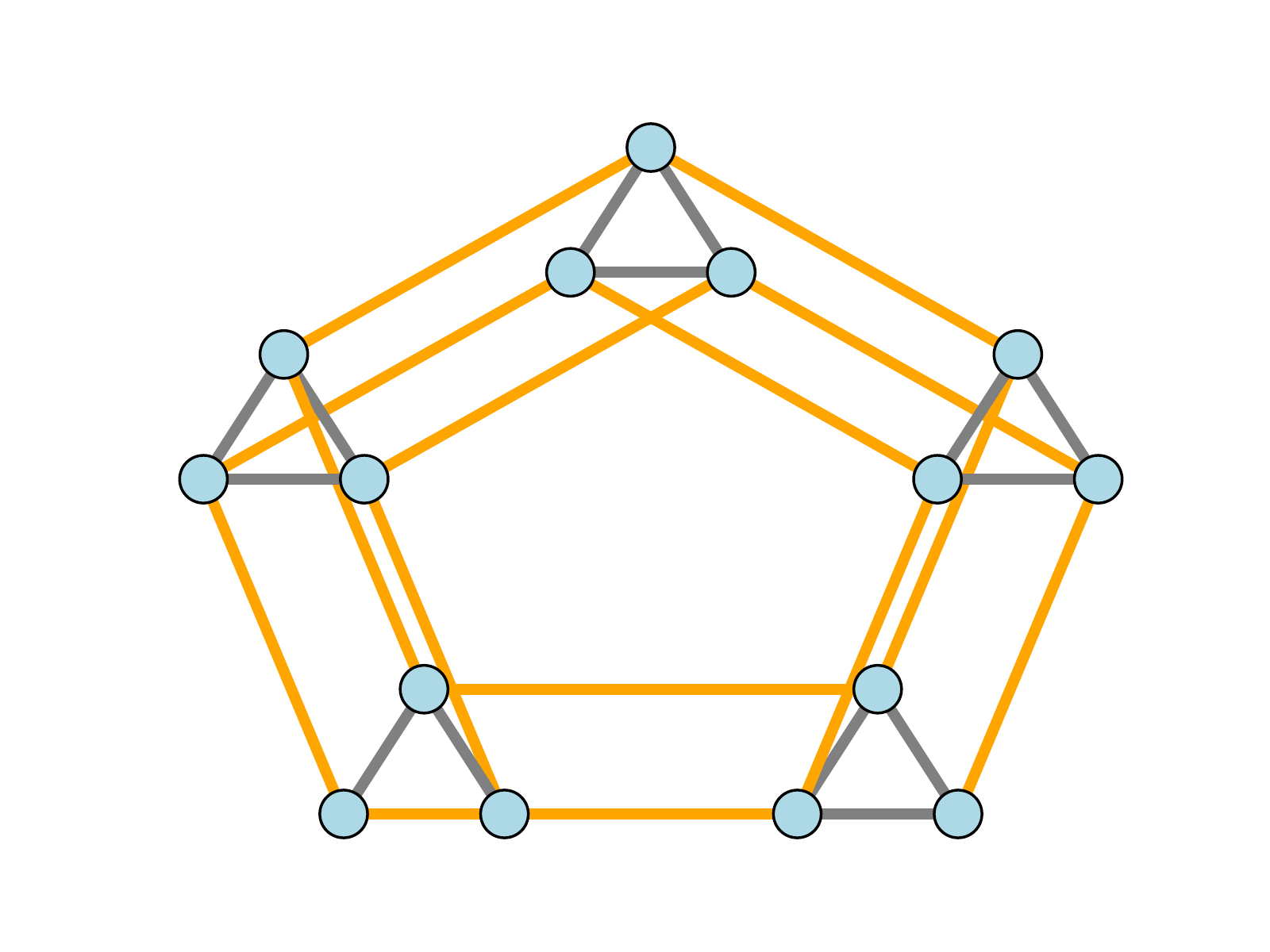}
  \put (48,5) {$\lambda_l$}
  \end{overpic}
  \end{subfigure}%
  \begin{subfigure}[t]{0.015\textwidth}
  \vspace{-5cm}
  {\myfont \Large C}
\end{subfigure}
  \adjustbox{minipage=1em}{\label{sfig:testb}}%
  \begin{subfigure}[t]{\dimexpr.36\linewidth}
  \centering
  \includegraphics[trim = 0 0 0 0, scale=.43]{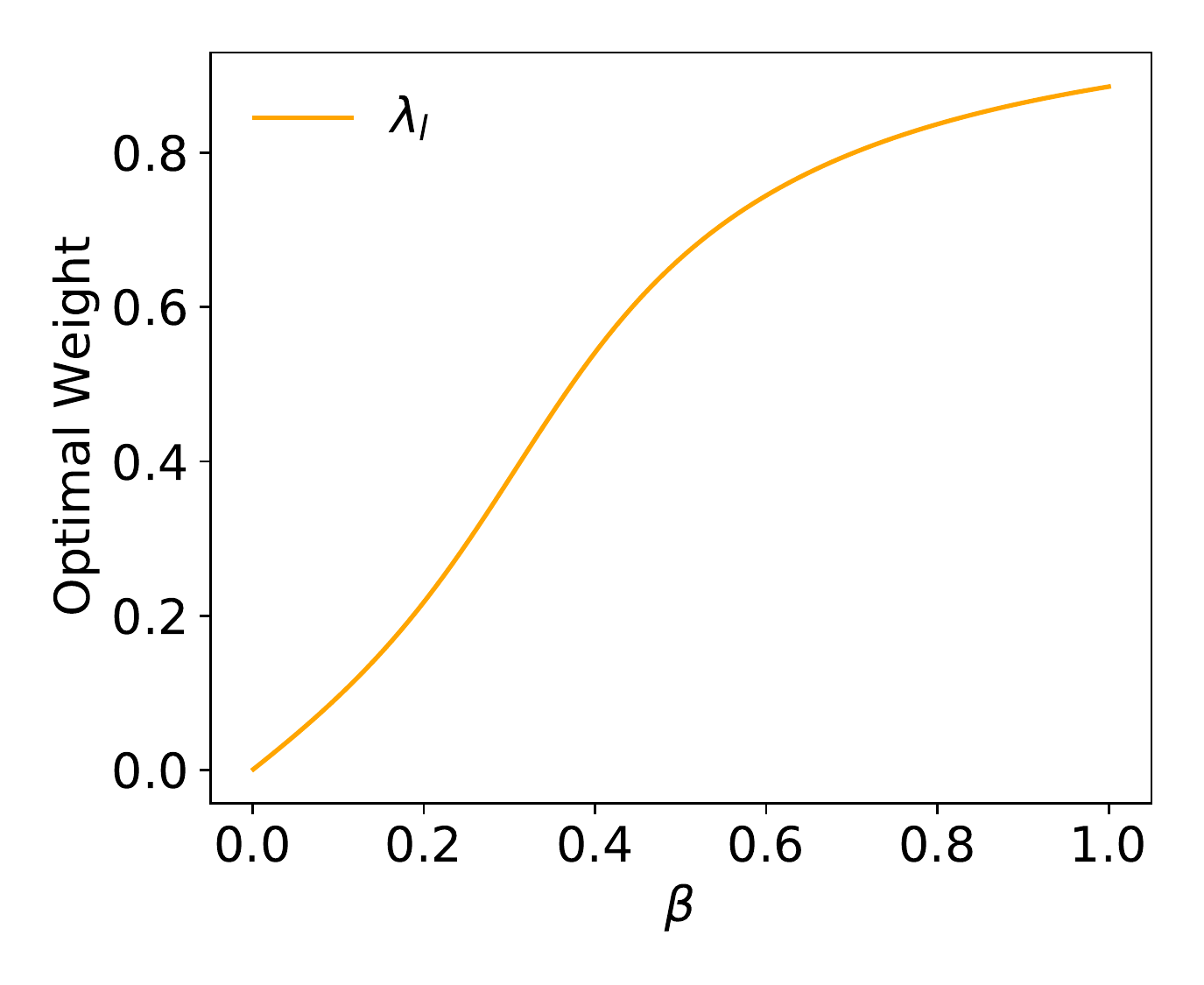}
  \end{subfigure}
  \\
  \vspace{-1cm}
  
  \begin{subfigure}[t]{0.015\textwidth}
  \vspace{.5cm}
  {\myfont \Large B}
\end{subfigure}
  \adjustbox{minipage=1em}{\label{sfig:testa}}%
  \begin{subfigure}[t]{\dimexpr.36\linewidth}
  \vspace{1em}
  \centering
        \includegraphics[trim = 70 0 0 0, clip, scale=.45]{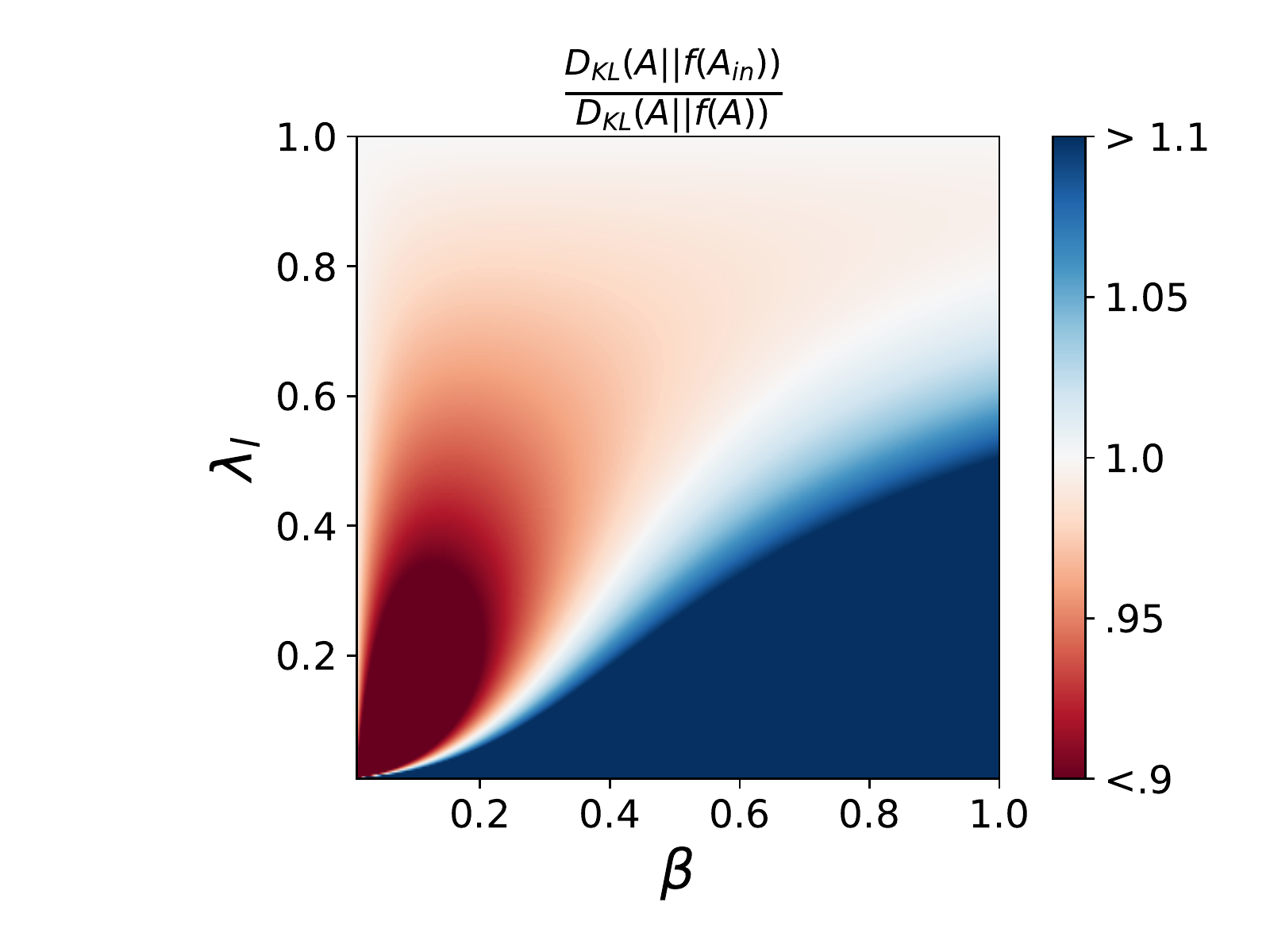}
  \end{subfigure}%
  \begin{subfigure}[t]{0.015\textwidth}
  \vspace{.5cm}
 {\myfont \Large D}
\end{subfigure}
\adjustbox{minipage=1em}{\label{sfig:testa}}%
  \begin{subfigure}[t]{\dimexpr.36\linewidth}
  \vspace{2em}
  \centering
      \includegraphics[trim = 0 0 0 0, clip, scale=.43]{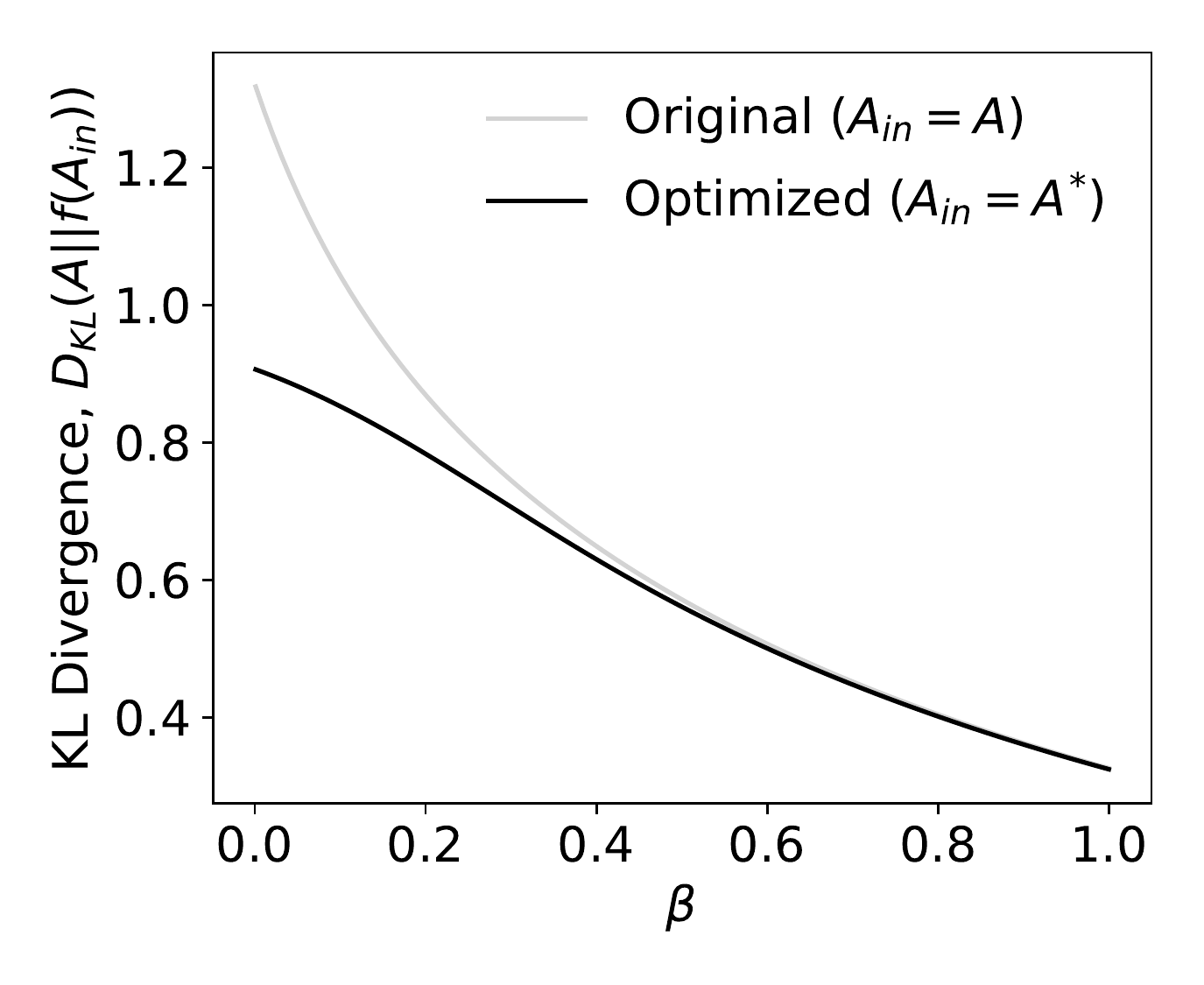}
  \end{subfigure}%
      \caption{\textbf{Optimizing the learnability of a lattice network.} \emph{(A)} A lattice network with $15$ nodes, each with degree $k_{i} = 4$, resulting in $30$ edges. \emph{(B)} Here we show the Kullback-Leibler divergence ratio (less than $1$ indicates enhanced learnability) across a section of the $\lambda_{l}$, $\beta$ parameter space. \emph{(C)} The optimal edge weight $\lambda_{l}$ for $0 < \beta < 1$. \emph{(D)} The Kullback-Leibler divergence between the learned network and the true network for different values of $\beta$, with and without input network optimization.}
  \label{fig:1}
\end{figure}

\begin{figure}
\begin{subfigure}[t]{0.015\textwidth}
  \vspace{-5cm}
  {\myfont \Large A}
\end{subfigure}
  \adjustbox{minipage=1em}{\label{sfig:testa}}%
  \begin{subfigure}[t]{\dimexpr.31\linewidth}
  \centering
      \includegraphics[trim = 50 0 0 0, clip, scale=.4]{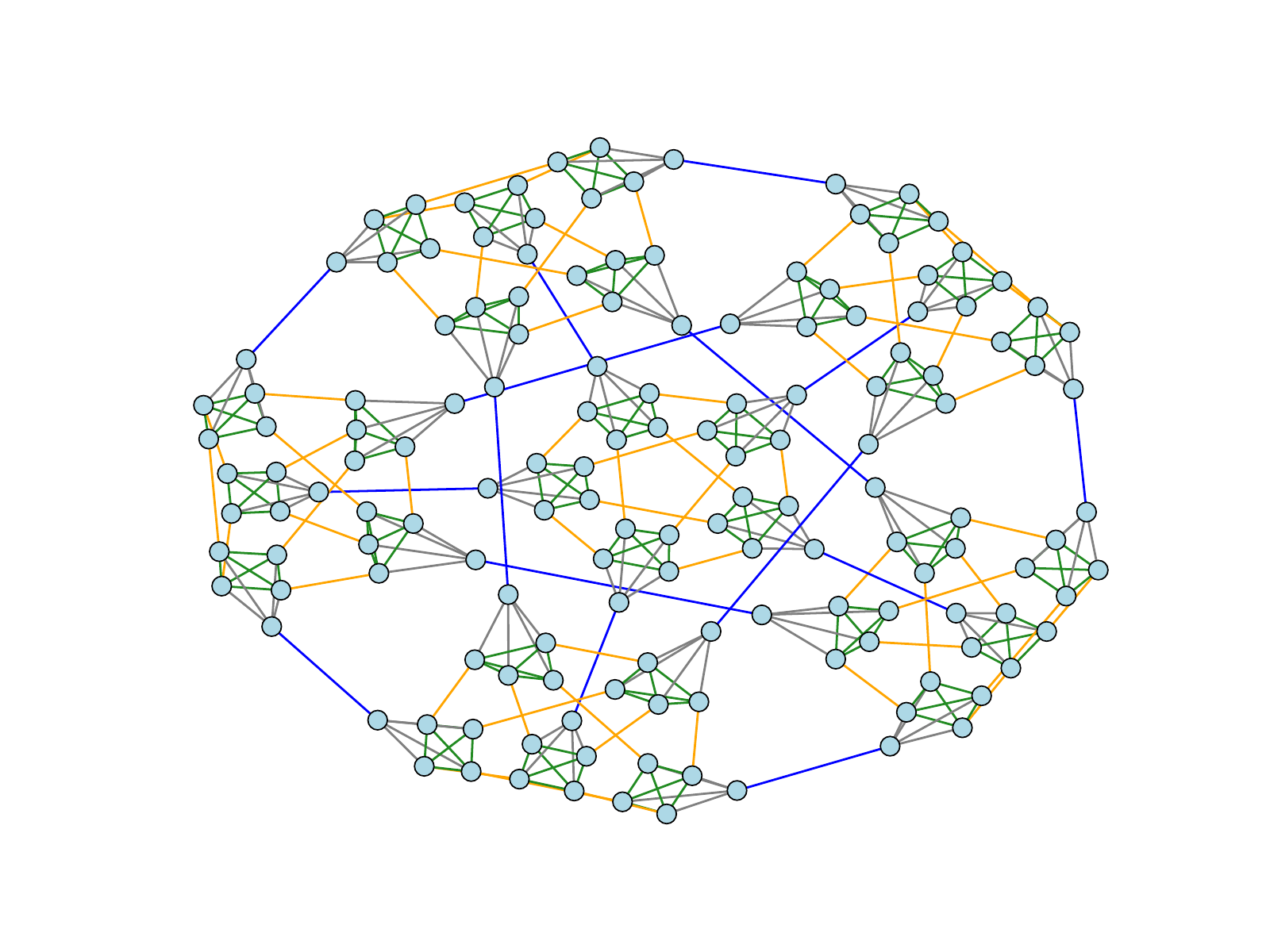}
  \end{subfigure}%
  \begin{subfigure}[t]{0.015\textwidth}
  \vspace{-5cm}
  {\myfont \Large C}
\end{subfigure}
  \adjustbox{minipage=1em}{\label{sfig:testb}}%
  \begin{subfigure}[t]{\dimexpr.31\linewidth}
  \centering
        \includegraphics[trim = 70 0 0 0, clip, scale=.4]{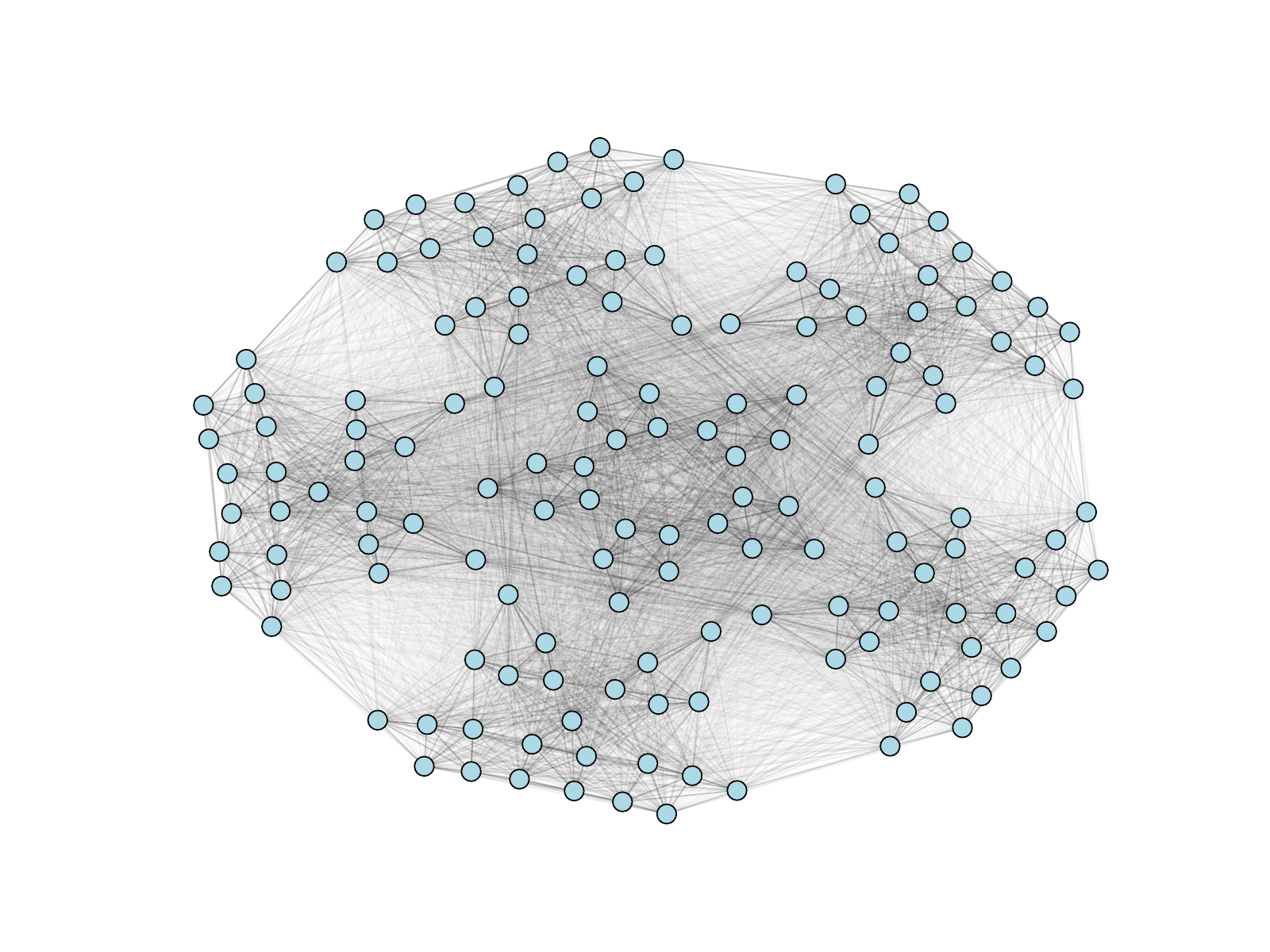}
  \end{subfigure}
  \begin{subfigure}[t]{0.015\textwidth}
  \vspace{-5cm}
  {\myfont \Large E}
\end{subfigure}
  \adjustbox{minipage=1em}{\label{sfig:testa}}%
  \begin{subfigure}[t]{\dimexpr.31\linewidth}
  \centering 
      \includegraphics[trim = 60 0 0 0, clip, scale=.4]{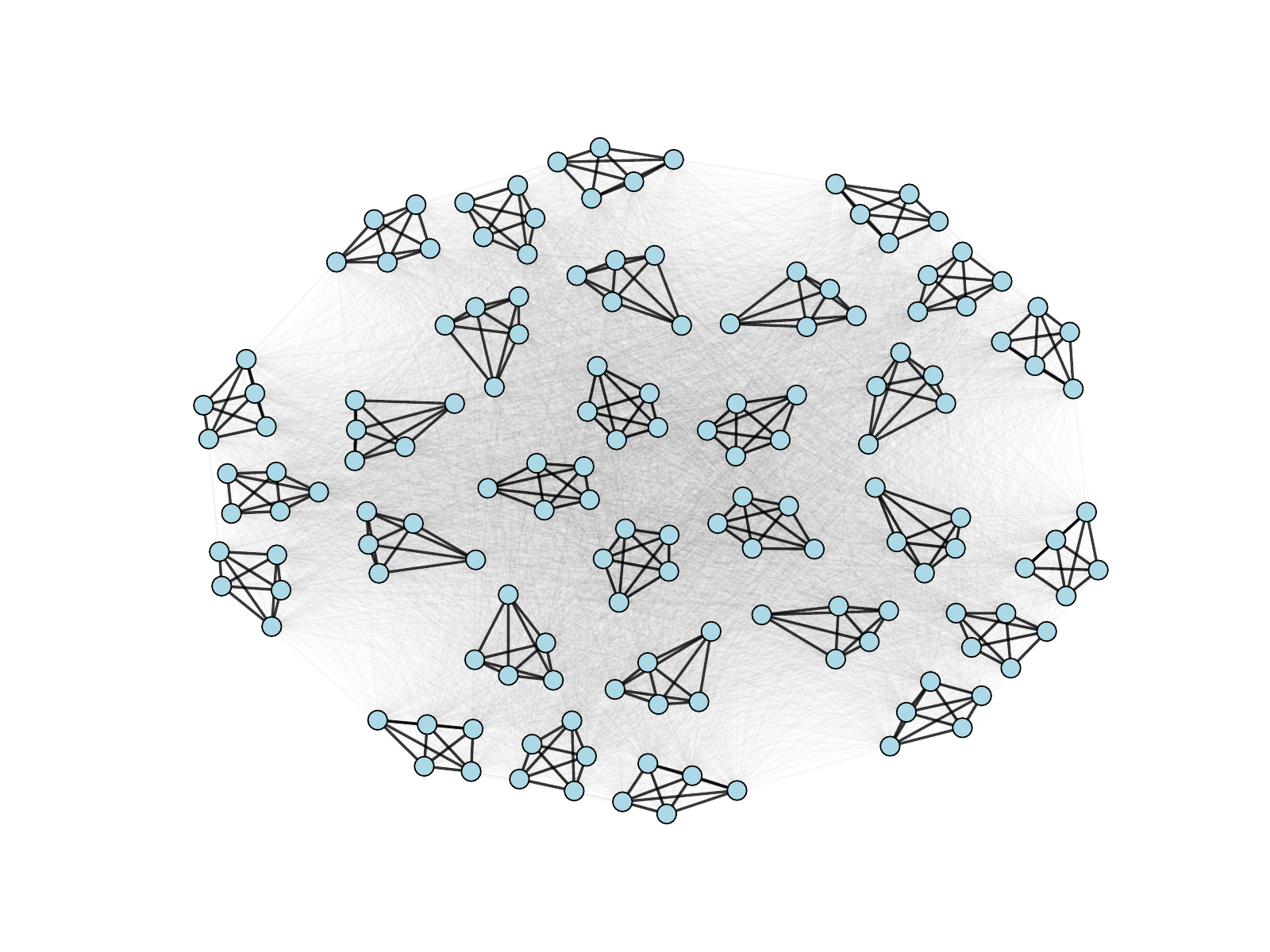}
  \end{subfigure}%
  \\
  \vspace{-1.5cm}
  
  \begin{subfigure}[t]{0.015\textwidth}
  \vspace{.5cm}
  {\myfont \Large B}
\end{subfigure}
  \adjustbox{minipage=1em}{\label{sfig:testa}}%
  \begin{subfigure}[t]{\dimexpr.32\linewidth}
  \vspace{2.45em}
  \centering
  \includegraphics[trim = 30 0 0 0, scale=.4]{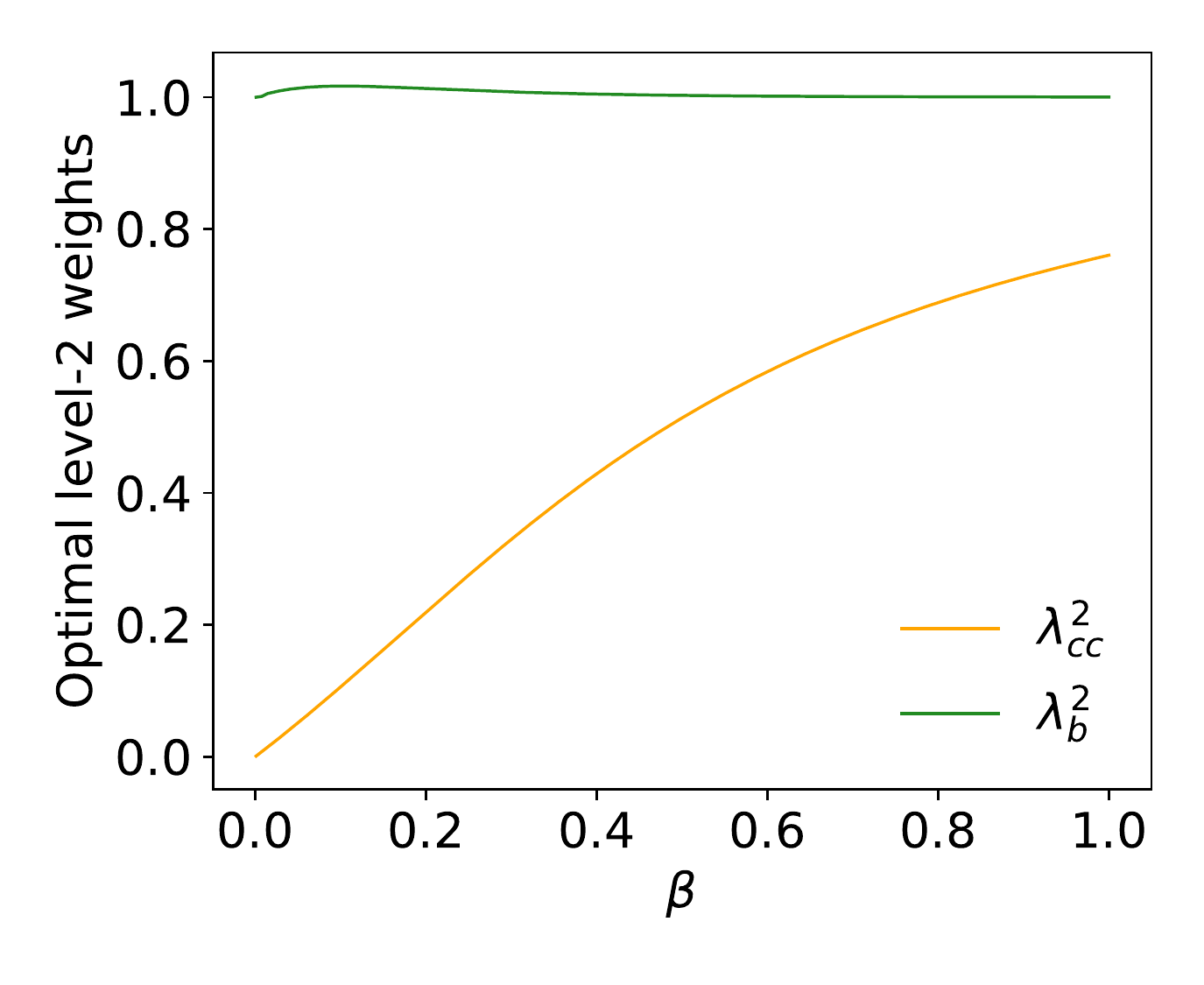}
  \end{subfigure}%
  \begin{subfigure}[t]{0.015\textwidth}
  \vspace{.5cm}
 {\myfont \Large D}
\end{subfigure}
\adjustbox{minipage=1em}{\label{sfig:testa}}%
  \begin{subfigure}[t]{\dimexpr.325\linewidth}
  \vspace{2.2em}
  \centering
      \includegraphics[trim = 15 0 0 10,clip, scale=.42]{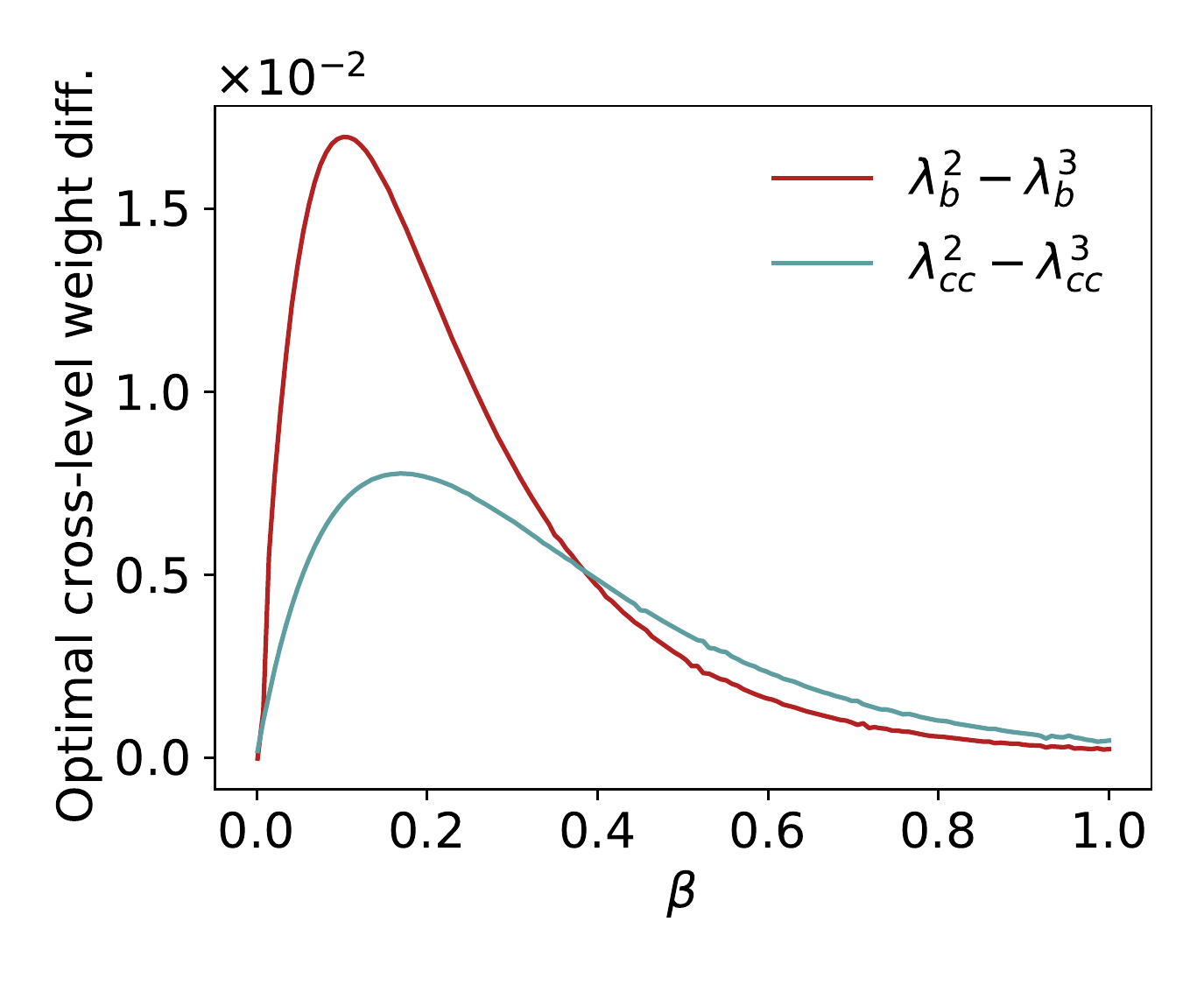}
  \end{subfigure}%
  \begin{subfigure}[t]{0.015\textwidth}
  \vspace{.5cm}
 {\myfont \Large F}
\end{subfigure}
\adjustbox{minipage=1em}{\label{sfig:testa}}%
  \begin{subfigure}[t]{\dimexpr.32\linewidth}
  \vspace{2.5em}
  \centering
      \includegraphics[trim = 30 0 0 0, scale=.4]{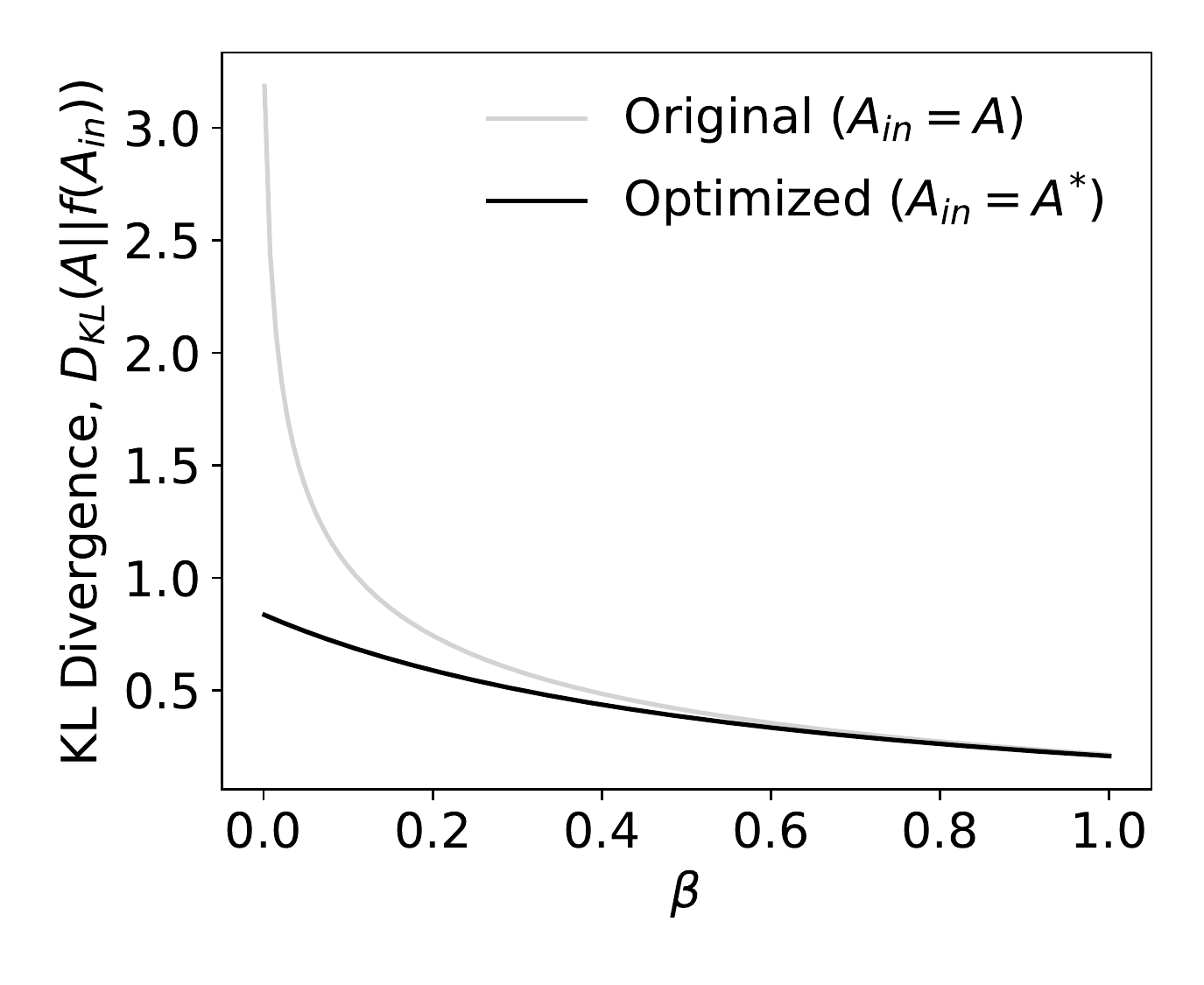}
  \end{subfigure}%
      \caption{\textbf{Optimizing the learnability of a Sierpiński network.} \emph{(A)} The Sierpiński network $S^{3}_5$ with $3$ levels, modified to have $6$ communities at the final level. \emph{(C, E)} The learned representations of the Sierpiński network at $\beta = 0.05$, both with \emph{(E)} and without \emph{(C)} input network optimization. \emph{(B)} The optimal level-2 edge weights $\lambda_{cc}^{2}$ and $\lambda_{b}^{2}$ for $0 < \beta < 1$. \emph{(D)} The differences $\lambda_{b}^{2} - \lambda _{b}^3$ and  $\lambda_{cc}^{2} - \lambda _{cc}^3$ between optimal edge weights of levels $2$ and $3$, for $0 < \beta < 1$. \emph{(F)} The Kullback-Leibler divergence between the learned network and the true network for different values of $\beta$, with and without input network optimization.}
  \label{fig:2}
\end{figure}

\begin{figure}
\centering
\begin{subfigure}[t]{0.015\textwidth}
  \vspace{-5cm}
  {\myfont \Large A}
\end{subfigure}
  \adjustbox{minipage=1em}{\label{sfig:testa}}%
  \begin{subfigure}[t]{\dimexpr.36\linewidth}
  \centering
  \vspace{-4.8cm}
      \includegraphics[trim = 0 0 0 0, clip, scale=.45]{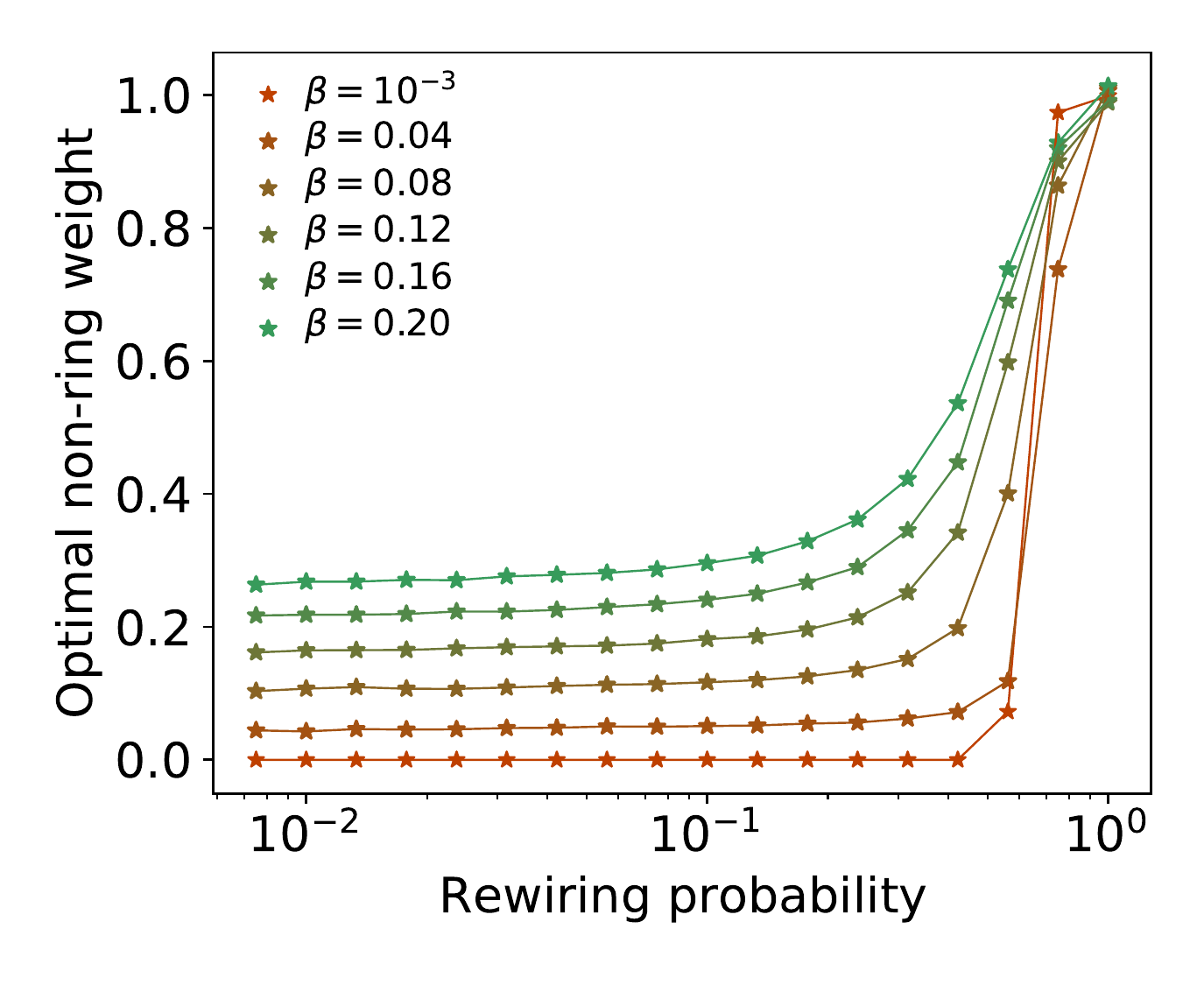}
  \end{subfigure}%
  \begin{subfigure}[t]{0.015\textwidth}
  \vspace{-5cm}
  {\myfont \Large B}
\end{subfigure}
  \adjustbox{minipage=1em}{\label{sfig:testb}}%
  \begin{subfigure}[t]{\dimexpr.36\linewidth}
  \centering
  \includegraphics[trim = 0 14 0 0, scale=.45]{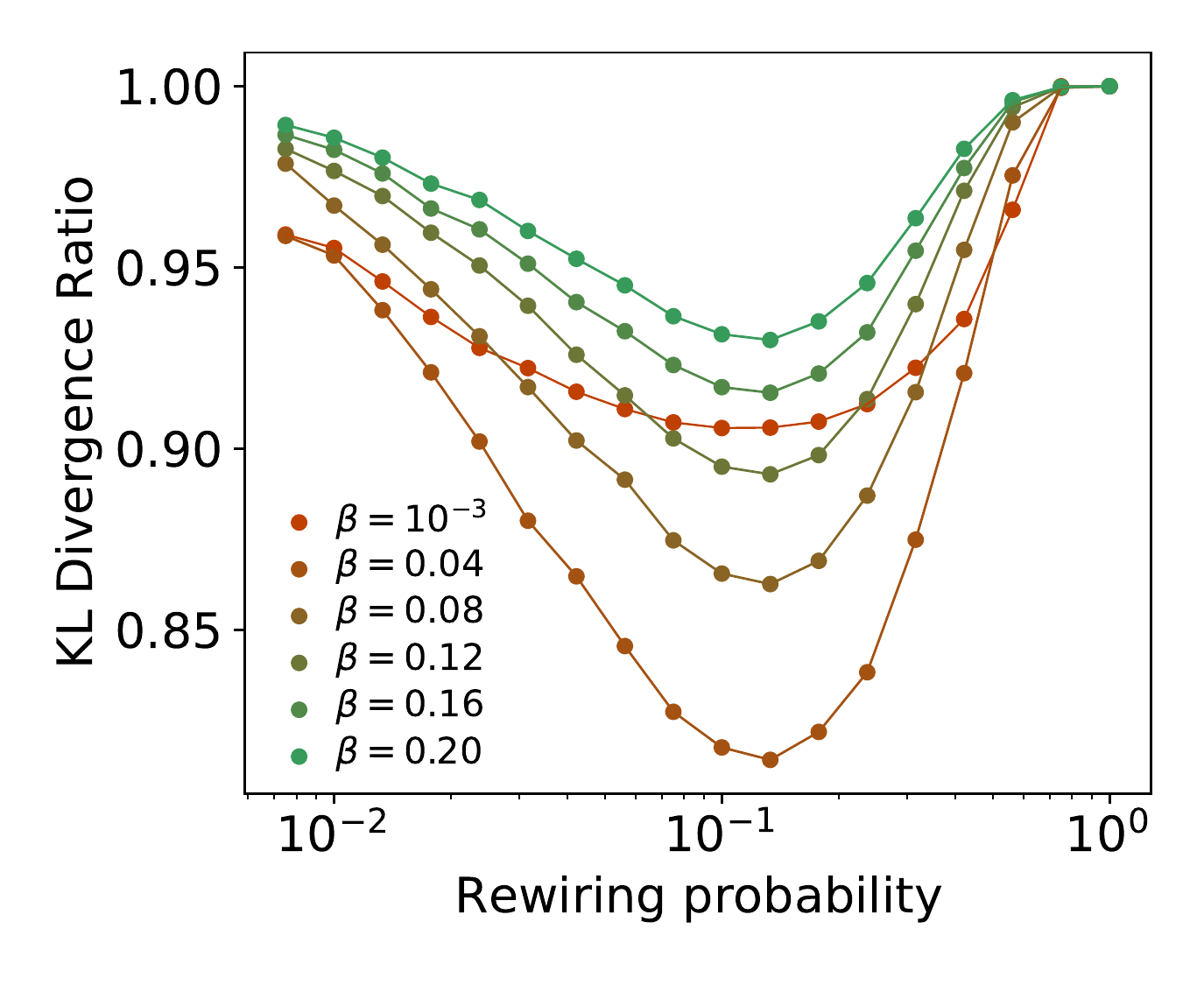}
  \end{subfigure}
  \\
 \vspace{-.8cm} 
  \begin{subfigure}[t]{0.015\textwidth}
  \vspace{.5cm}
  {\myfont \Large C}
\end{subfigure}
  \adjustbox{minipage=1em}{\label{sfig:testa}}%
  \begin{subfigure}[t]{\dimexpr.70\linewidth}
  \vspace{3em}
  \centering
  \begin{overpic}[width=.9\textwidth,tics=10]{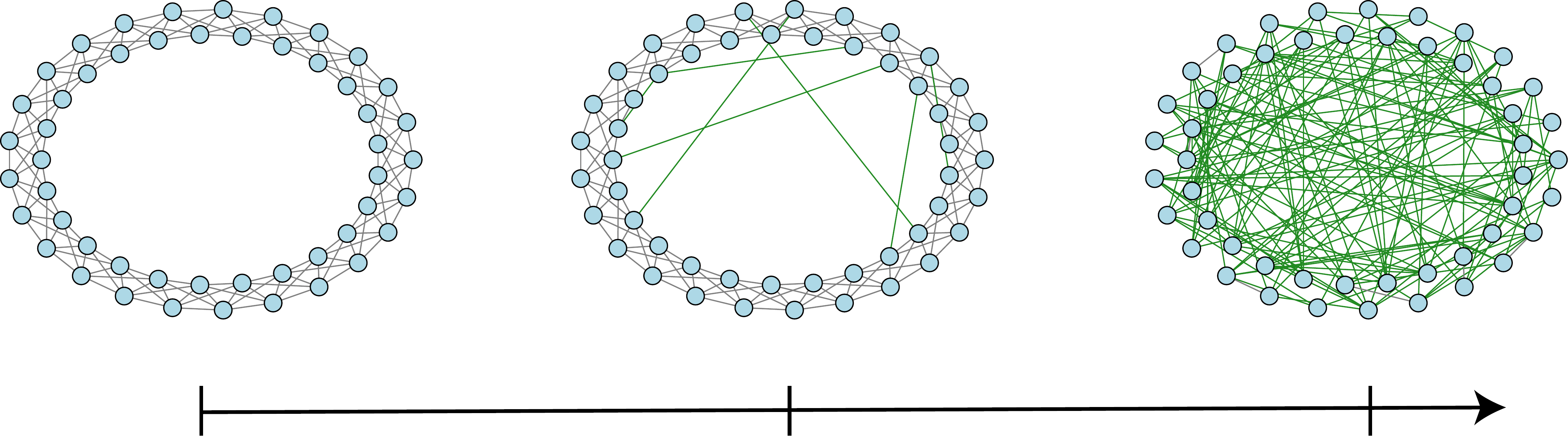}
  \put(12, -3){$0$}
  \put(49, -3){$10^{-1}$}
  \put(86, -3){$10^{0}$}
  \put(98, 1){\Large $p$}
\end{overpic}
  \end{subfigure}%
  \vspace{1em}
    \caption{\textbf{Optimizing the learnability of small world networks.} \emph{(A)} The optimal non-ring edge weight $\lambda_{nr}$ for enhancing learnability versus the rewiring probability $p$ of a Watts-Strogatz network at different values of $\beta$. \emph{(B)} The Kullback-Leibler divergence ratio $\frac{D_{KL}(A||f(A_{\text{in}}))}{D_{KL}(A||f(A))}$ achieved with optimal non-ring edge weights at different values of $\beta$. The findings reported in panels \emph{(A,B)} represent results obtained for networks with $N = 200$ nodes and an average degree of $\langle k \rangle = 10$. Each curve is an average over the results from $25$ generated networks. \emph{(C)} A schematic demonstrating how the structure of Watts-Strogatz networks changes as the rewiring probability $p$ increases. Non-ring edges are shown in green.}
      \label{fig:3}
\end{figure}

\begin{figure}
\begin{subfigure}[t]{\dimexpr\linewidth}
    \centering
    \begin{overpic}[width=.55\textwidth,tics=10]{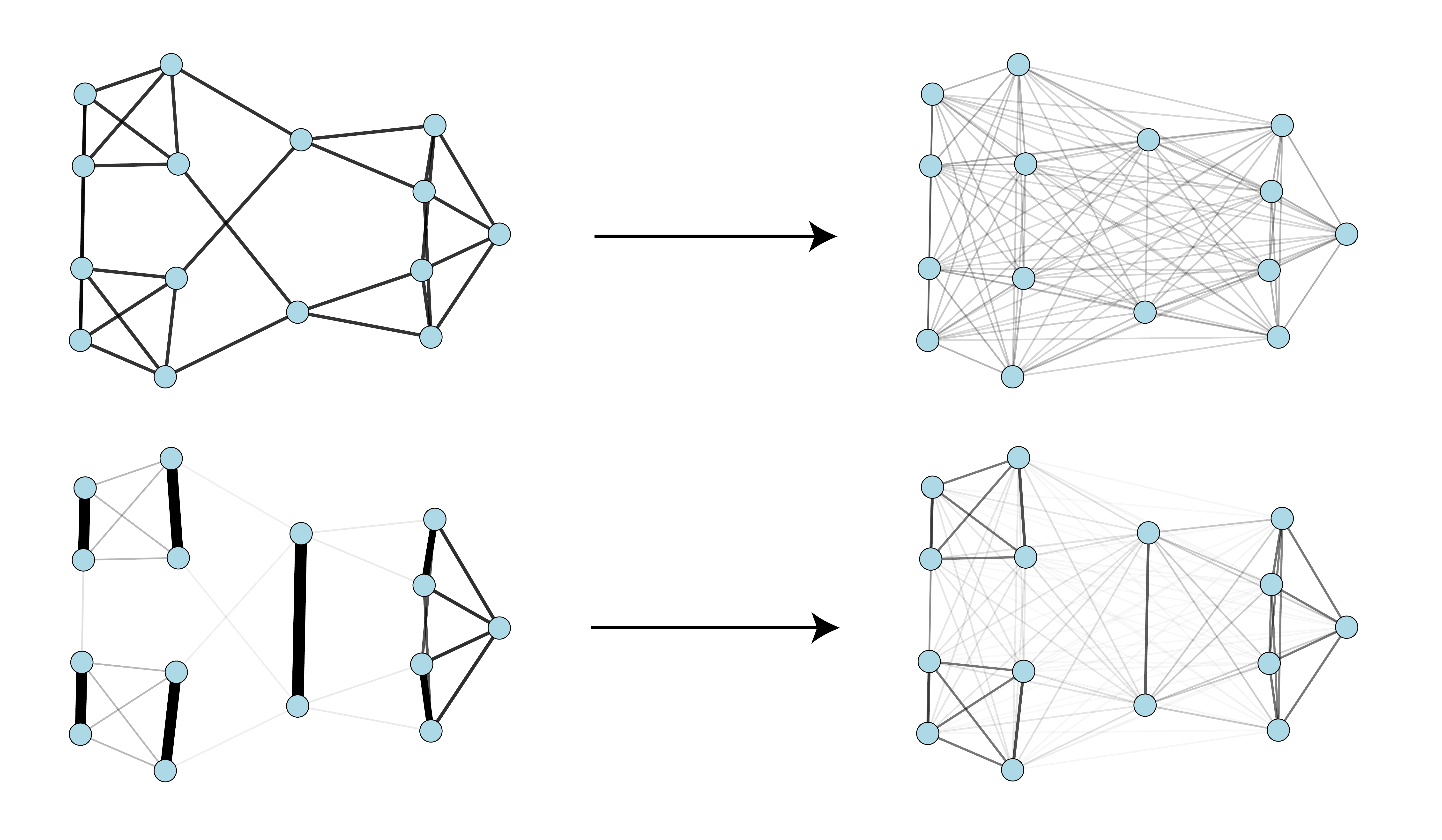}
 \put (36,50) {\Large$\displaystyle A$}
 \put (36, 18.5) {\Large$\displaystyle A^{*}$}
 \put (97, 50) {\Large$\displaystyle f(A)$}
 \put (97, 18.5) {\Large$\displaystyle f(A^{*})$}
\end{overpic}
\end{subfigure}
    \caption{\textbf{Optimal emphasis modulation of an example network when considering nonexistent edges.} Here we show the learned networks resulting from human learning of an example network (\emph{top}), as well as from the example network optimized for learnability (\emph{bottom}). The optimized network was determined with the addition of nonexistent edges as free parameters. Optimized and learned networks were both computed at $\beta = 0.05$. }
    \label{fig:4}
\end{figure}

\paragraph{Watts-Strogatz networks.} Our analysis of the lattice network (Fig. \ref{fig:1}\emph{A}) demonstrated that edges that do not contribute to the formation of small clusters or triangles should be de-emphasized in order to optimize network learnability. To assess this conclusion in a more general class of networks, we consider the optimization of learnability for Watts-Strogatz networks. Prior to any rewiring ($p = 0$), such networks begin as a ring-like lattice of nodes, with each node only having connections to its nearest neighbors in the ring (Fig. \ref{fig:3}\emph{C}). Given the density of local connections, these ring-like networks are highly clustered. When a small fraction of edges are then rewired, Watts-Strogatz networks maintain similar levels of clustering, but display markedly lower average path lengths \cite{Humphries_Gurney_2008, Latora_Marchiori_2001}. In this regime, Watts-Strogatz networks can often be characterized by small-worldness, a concept relevant to a number of real-world networks including brain networks, language networks, and metabolic networks \cite{Bassett_Bullmore_2006, Cancho_Sole_2001, Wagner_Fell_2001, Telesford_Joyce_Hayasaka_Burdette_Laurienti_2011, de_Arcangelis_Herrmann_2002}. Finally, in the limit of high rewiring $p = 1$, the structure of Watts-Strogatz networks is very similar to that of Erdős–Rényi networks. Motivated by previous work reporting that networks with high clustering coefficients are more learnable \cite{Lynn_Papadopoulos_Kahn_Bassett_2020}, we investigate the optimization of learnability in Watts-Strogatz networks at different rewiring probabilities $p$. In particular, we seek to determine whether the rewired edges, which deviate from the original highly-clustered ring network, ought to be de-emphasized when presented to human learners. In addition, we aim to identify whether the efficacy of strategies for optimizing network learnability depend on the emergence of small-world structure, which tends to appear for rewiring probabilities of $10^{-2} \le p \le 10^{-1}$ \cite{Watts_Strogatz_1998}. 

To investigate how rewired edges in Watts-Strogatz networks should be weighted to maximize network learnability, we consider the optimal weight $\lambda_{nr}$ of rewired edges relative to non-rewired edges on the ring. For low rewiring probabilities ($p < 10^{-0.5}$), we find that network learnability is optimized by de-emphasizing rewired edges and over-emphasizing edges on the ring (Fig. \ref{fig:3}\emph{A}). Considering that the original lattice-like ring is highly clustered, and is therefore naturally easier to learn \cite{Lynn_Papadopoulos_Kahn_Bassett_2020}, these findings suggest that de-emphasizing areas of a network that do not contribute to clustering may be an effective general strategy for enhancing network learnability. This finding is consistent over the range $ 10^{-3} \leq \beta \leq 0.2$ of $\beta$ values analyzed. In particular, in the limit $\beta \rightarrow 0$, the optimal non-ring edge weight approaches $\lambda_{nr} \rightarrow 0$ for nearly all rewiring probabilities $p$, whereas higher $\beta$ values (more accurate learning) do not require such stark de-emphasis of non-ring edges. In addition, as the rewiring probability $p$ approaches $1$, the weight given to non-ring edges approaches $1$. Given that highly-rewired Watts-Strogatz networks are equivalent to random Erdős–Rényi networks, it is reasonable that for high values of $p$, there is no distinction between ring and non-ring edges. 

Interestingly, we also find that improvements in learnability resulting from tuning non-ring edge weights are most prominent at intermediate rewiring probabilities near $p \sim 10^{-1}$ (Fig \ref{fig:3}\emph{B}). This finding suggests that the learnability of networks with small-world structure is significantly more optimizable when compared to highly ordered lattice-like networks or to highly disordered Erdős–Rényi networks.

\clearpage
\bibliography{pnas-supplement}
\bibliographystyle{unsrt}